\documentclass[sigplan,10pt]{acmart}

\setcopyright{none}
\renewcommand\footnotetextcopyrightpermission[1]{}

\usepackage{enumitem}
\usepackage{amsmath}
\usepackage{algorithmic}
\usepackage{graphicx}
\usepackage{textcomp}
\usepackage[table]{xcolor}
\usepackage{tikz}
\usepackage{multirow}
\usepackage{booktabs}
\usepackage{pifont}
\usepackage{makecell}
\usepackage{caption}
\usepackage[most]{tcolorbox}
\usepackage{array}
\usepackage{booktabs}
\usepackage{makecell}
\usepackage{tabularx}
\definecolor{forestgreen}{RGB}{34,139,34}
   
\newcommand{\datasetrow}[5]{%
\textbf{#1}: #2\\[0.35em]
\hspace*{2em}#3\hfill #4\hfill #5
}
\usepackage{pgfplots}
\pgfplotsset{compat=1.18}
\definecolor{selectedrow}{RGB}{224,238,249}
\usepackage{subcaption}
\usetikzlibrary{positioning,fit,calc,backgrounds,arrows.meta}
\def\BibTeX{{\rm B\kern-.05em{\sc i\kern-.025em b}\kern-.08em
    T\kern-.1667em\lower.7ex\hbox{E}\kern-.125emX}}

\newcommand{\thiswork}{Falcons}

\setlist{
  ,nosep
  ,itemindent=0pt
  ,leftmargin=*
  ,before=\vspace{.25\baselineskip}
  ,after=\vspace{.25\baselineskip}
}

\title[\thiswork]{\thiswork{}: Pushing the Speed-Ratio Limits of Lossy Compression for Parallel Scientific Applications}

\begin{document}



\author{Bing Lu}
\orcid{0009-0007-7633-0486}
\affiliation{
  \institution{Institute of Computing Technology, Chinese Academy of Sciences}
  \city{Beijing}
  \country{China}
}
\affiliation{%
  \institution{University of Chinese Academy of Sciences}
  \city{Beijing}
  \country{China}
}
\email{lubing26@mails.ucas.ac.cn}

\author{Jinyang Liu}
\authornote{Corresponding author: Jinyang Liu.}
\affiliation{%
  \institution{Shanghai Jiao Tong University}
  \city{Shanghai}
  \country{China}
}
\email{jinyang.liu@sjtu.edu.cn}

\author{Xinyang Ma}
\affiliation{
  \institution{Institute of Computing Technology, Chinese Academy of Sciences}
  \city{Beijing}
  \country{China}
}
\affiliation{%
  \institution{University of Chinese Academy of Sciences}
  \city{Beijing}
  \country{China}
}
\email{maxinyang25@mails.ucas.ac.cn}

\author{Quantong Fu}
\affiliation{%
  \institution{Shanghai Jiao Tong University}
  \city{Shanghai}
  \country{China}
}
\email{quantongfu@sjtu.edu.cn}

\author{Chenwei Hou}
\affiliation{%
  \institution{Shanghai Jiao Tong University}
  \city{Shanghai}
  \country{China}
}
\email{moonlightplague@sjtu.edu.cn}

\author{Zedong Liu}
\affiliation{%
  \institution{Institute of Computing Technology, Chinese Academy of Sciences}
  \city{Beijing}
  \country{China}
}
\email{captain.liu77@gmail.com}

\author{Wenjing Huang}
\affiliation{%
  \institution{Institute of Computing Technology, Chinese Academy of Sciences}
  \city{Beijing}
  \country{China}
}
\email{huangwenjing23@mails.ucas.ac.cn}

\author{Yida Gu}
\affiliation{%
  \institution{Institute of Computing Technology, Chinese Academy of Sciences}
  \city{Beijing}
  \country{China}
}
\email{guyida21@mails.ucas.ac.cn}

\author{Xingchen Liu}
\affiliation{%
  \institution{Institute of Computing Technology, Chinese Academy of Sciences}
  \city{Beijing}
  \country{China}
}
\email{liuxingchen232@mails.ucas.ac.cn}

\author{Zheng Wei}
\affiliation{%
  \institution{Institute of Computing Technology, Chinese Academy of Sciences}
  \city{Beijing}
  \country{China}
}
\email{weizheng@ncic.ac.cn}

\author{Anbang Wu}
\affiliation{%
  \institution{Shanghai Jiao Tong University}
  \city{Shanghai}
  \country{China}
}
\email{anbang@cs.sjtu.edu.cn}

\author{Xiaofeng Hou}
\affiliation{%
  \institution{Shanghai Jiao Tong University}
  \city{Shanghai}
  \country{China}
}
\email{xfhelen@sjtu.edu.cn}

\author{Chao Li}
\affiliation{%
  \institution{Shanghai Jiao Tong University}
  \city{Shanghai}
  \country{China}
}
\email{chaol@sjtu.edu.cn}

\author{Guangming Tan}
\orcid{0000-0002-6361-5948}
\affiliation{%
  \institution{Institute of Computing Technology, Chinese Academy of Sciences}
  \city{Beijing}
  \country{China}
}
\email{tgm@ict.ac.cn}

\author{Minyi Guo}
\affiliation{%
  \institution{Shanghai Jiao Tong University}
  \city{Shanghai}
  \country{China}
}
\email{myguo@sjtu.edu.cn}

\author{Dingwen Tao}
\orcid{0000-0001-5422-4497}
\affiliation{%
  \institution{Institute of Computing Technology, Chinese Academy of Sciences}
  \city{Beijing}
  \country{China}
}
\email{taodingwen@ict.ac.cn}

\begin{abstract}
The extreme volume of scientific data has become a critical challenge for modern exa-scale scientific computing applications,  and error-bounded lossy compression remains indispensable for scientific data management due to its high data reduction rate and controllable data fidelity. However, to ensure sufficient data reduction, existing compressors incur prohibitive computational overheads in real-world parallel scientific computing applications. To overcome the performance bottlenecks of error-bounded lossy compression, we present \textbf{\thiswork{}}, a broadly deployable error-bounded lossy compressor that simultaneously achieves \textbf{high compression ratios with ultra-fast throughput,  minimizing the overhead in parallel scientific applications}. \thiswork{} integrates three key designs: memory-regularized interpolation with SIMD (Single Instruction Multiple Data) vectorization, distribution-adaptive lossless coding, and application-oriented optimizations for multiple parallelization schemes.
Across our evaluations on cutting-edge supercomputers, 
\thiswork{} clearly outperforms state-of-the-art compressors, simultaneously achieving up to 29\% compression ratio improvement and 255\% throughput improvement. In case studies of real-world applications, \thiswork{} accelerates large-scale parallel data I/O and remote transfer by up to 44.6\% and 51.5\%, and provides 1.55$\times$ end-to-end speedup for the full WRF (Weather Research \& Forecasting) application workflow.
\end{abstract}
\maketitle
\pagestyle{plain}
\thispagestyle{plain}

\section{Introduction}
\label{sec:intro}

Advances in supercomputers, large-scale scientific instruments, and AI4S models dramatically drive progress in scientific discovery. However, the explosion of data volumes in turn severely challenges the data storage and management systems.
For example, although a $16384^3$-element NYX~\cite{almgren2013nyx,lukic2015lyman,sexton2021nyx} cosmology simulation is computationally feasible on OLCF’s Frontier, at least 14 arrays per checkpoint and 20 snapshots per science case require nearly 10 PB of storage, making data storage and transfer a major bottleneck.


Scientific data-management workflows now span a spectrum of computing facilities, ranging from personal workstations to CPU-centric leadership supercomputers (e.g., Fugaku and LineShine~\cite{fugaku,lineshine}) and heterogeneous exascale systems (e.g., Aurora and El Capitan~\cite{alcf2024aurora,elcapitan}). Across these platforms, applications call for effective data-reduction solutions to alleviate growing bottlenecks in storage, parallel I/O, and cross-site data transfer. Error-bounded lossy compression~\cite{lindstrom2014fixed,lu2023zfp,tao2017significantly,di2016fast,Xin-bigdata18,szinterp,sz3,liu2022dynamic,liu2024high,sperr,ballester2019tthresh}, which guarantees a strict pointwise error bound, has been widely adopted for scientific data management~\cite{cappello2019use,grosset2020foresight,cappello2025multifacets,2025survey,sc25qoi}, as it substantially reduces data volume while preserving the fidelity required for downstream analysis. While compression can also be offloaded to accelerators such as GPUs~\cite{cuszhi,cuszp3,lindstrom2014fixed}, CPU-based compression remains indispensable: many leadership systems and the workstations where scientific data are prepared and analyzed have no accelerator available for this purpose, and CPU implementations integrate directly into existing scientific software stacks and system services~\cite{kay2015community,xu2025kilometer,skamarock2019description,hdf5} while remaining portable across platforms.

\begin{table}[t]
\centering
\footnotesize
\setlength{\tabcolsep}{3pt}
\renewcommand{\arraystretch}{1}
\caption{Breakdown of parallel writing time (s) on the LineShine supercomputer with 64 nodes and 4096 MPI ranks. Dataset: Hurricane Isabel~\cite{hurricane}. Error Bound: $0.1\%$ of the value range. Time\_C is for compression, and Time\_W is for data writing. Only \thiswork{} breaks the bottleneck for compression execution.}
\label{tab:hurricane_pio_4096}

\resizebox{0.9\columnwidth}{!}{%
\begin{tabular}{@{}c c c >{\columncolor{gray!15}}c >{\columncolor{gray!15}}c@{}}
\toprule
Compressor & Time\_C & Time\_W & Time\_Total & Comp\_Ratio \\
\midrule
Original I/O & / & 16.30 & 16.30 & 1.00$\times$ \\
ZFP~\cite{lindstrom2014fixed} & 12.00 & 6.39 & 18.39 & 6.68$\times$ \\
SZ3.3~\cite{sz3} & 24.12 & 4.30 & 28.42 & 23.18$\times$ \\
SPERR~\cite{sperr} & 78.86 & 4.13 & 82.99 & 15.19$\times$ \\
\textbf{\thiswork{} (ours)} & 4.51 & 4.10 & \textbf{\underline{8.61}} & \textbf{\underline{26.07$\times$}} \\
\bottomrule
\end{tabular}%
}
\end{table}

So far, the performance of scientific lossy compressors in real-world parallel applications is bottlenecked by \textbf{the dilemma of compression ratio and throughput}.
To achieve sufficient data reduction, lossy compression algorithms are computationally intensive and incur non-negligible performance overheads~\cite{chen2025computational}. 
A typical use case of lossy compression in scientific applications is to accelerate data I/O~\cite{sc25qoi,cappello2025lossy,Xin-bigdata18,szinterp}. However, 
as indicated in Table~\ref{tab:hurricane_pio_4096}, which profiles large-scale parallel data writing on the newly released LineShine Supercomputer, which has $>$ 2 ExaFlops peak performance and leading I/O bandwidth,
existing scientific compressors even slow down the I/O workflow in certain cases, because their execution time emerges as a bottleneck when reducing the data volume to an acceptable scale. 
Thus, scientific lossy compression must be comprehensively redesigned to achieve a joint optimization of compression ratio and throughput, fully supporting modern applications.

Having overcome the challenges of \textbf{performing effective scientific data decorrelation/encoding at minimized cost}, and \textbf{fine-tuning their performance in sophisticated real-world applications}, we propose \textbf{\thiswork{}}\footnote{We open-source the code at \url{https://github.com/opensdc-org/Falcons.git}.}. ~\thiswork{} is a prediction-based scientific error-bounded lossy compressor facilitating efficiency-boosted data interpolation design, adaptive lossless encoding modules, and parallelization-oriented optimizations. 
\thiswork{} is a CPU-based scientific lossy compressor to simultaneously deliver ultra-fast throughput and high compression ratios across parallel schemes, empowering scientific data applications to maximize their performance in diverse use cases.
Our contributions are summarized as follows:

\begin{itemize}
\item We design a novel interpolation-based data prediction scheme that exposes regularized memory access and is highly vectorizable on modern CPU architectures. It can leverage AVX2 and SVE2 intrinsics to fully unlock the computational capabilities of modern x86/ARM CPUs.

\item We design an adaptive entropy-coding pipeline tailored to the statistical characteristics of post-prediction quantization codes to fit the scientific data inputs, significantly outperforming existing scientific lossless encoding pipelines.

\item We further optimize \thiswork{} across different parallelism in real-world parallel applications. We reduce memory traffic to mitigate cache/memory bandwidth competitions and exploit fine-grained workload partition to preserve high throughput while maintaining compression effectiveness.


\item We conduct comprehensive evaluations for \thiswork{} on cutting-edge supercomputers.
It significantly outperforms all existing baselines, achieving both compression ratio improvement (up to 29\%) and speedup (up to 255\%). \thiswork{} boosts large-scale parallel data I/O and remote transfer by 44.6\% and 51.5\% time reduction, and enables 1.55$\times$ end-to-end application workflow speedup.
\end{itemize}

The rest of this paper is organized as follows. Section~\ref{sec:rw} discusses the related work. Section~\ref {sec:rb} presents the background and formulation of our research. In Section~\ref{sec:overview} and Section~\ref {sec:design}, we present the overview and details of the \thiswork{} design. Section~\ref{sec:eva} demonstrates the evaluation results. Finally, we present the concluding remark and future work in Section~\ref{sec:con}.
\section{Related Work}
\label{sec:rw}



\subsection{Scientific Error-Bounded Lossy Compression}
Various error-bounded lossy compressors have been developed for scientific data reduction to support scientific applications. Existing works can be broadly categorized into four classes: prediction-based, transform-based, dimension-reduction-based, and neural-network-based. 
The SZ family is a representative prediction-based framework, employing a variety of predictors, such as the mixed scheme of Regression-Lorenzo prediction in SZ2~\cite{Xin-bigdata18} and dynamic spline interpolation in SZ3~\cite{szinterp,sz3,liu2022dynamic}.
 MGARD~\cite{MGARD} uses an error-guaranteed theoretical framework of multi-level decomposition. ZFP~\cite{lindstrom2014fixed,zfp2025} delivers fast throughput by using a near-orthogonal transform for data decorrelation
. SPERR~\cite{sperr} achieves superior rate–distortion performance but low throughput by utilizing the CDF 9/7 discrete wavelet transform. 
Dimension-reduction-based compressors apply dimension-reduction techniques, such as high-order singular value decomposition (HoSVD) in TTHRESH~\cite{ballester2019tthresh}, which achieves high compression ratios but very low throughput.
Neural-network-based scientific data compressors~\cite{ae-sz,SRNN-SZ,DCINR,jia2024gwlz,sanjaysc25} exploit learned data representations with significantly low speed, limiting their availability in real systems.

\subsection{Accelerating Scientific Lossy Compression in Applications}
To optimize the efficiency of the scientific data management workflow with scientific lossy compression, numerous studies have been proposed. PFPL~\cite{pfpl} targets heterogeneous HPC workflows by providing bitwise-compatible compression, enabling data to be compressed and decompressed on different devices while maintaining high throughput. SZ3\_SIMD~\cite{sz3simd} introduces SIMD-based optimization of the Lorenzo data predictor~\cite{fpzip}, thereby achieving good speed but a poor compression ratio. MGARD+~\cite{liang2021mgard} enhances the MGARD~\cite{MGARD} framework with level-wise quantization and adaptive decomposition, improving compression ratio and throughput.
Several studies have also explored GPU-accelerated scientific data compression~\cite{tian2020cusz,cusz+,liu2024cusz,cuszhi,huang2023cuszp,huang2024cuszp2,cuszp3,lu2026prism,zhang2023fz}, but they generally achieve lower compression ratios than CPU-based compressors. 

In summary, existing error-bounded lossy compressors have not achieved jointly optimized compression ratio, throughput, and scalability, exhibiting limitations on universal integration in parallel scientific applications. Our paper endeavors to make a breakthrough for this research gap. 


\section{Research Background}
\label{sec:rb}

    \subsection{The Principle and Use Cases of Error-Bounded Lossy Compression}
\label{sec:bg:data_management}

\begin{figure}[t]
    \centering
    \includegraphics[width=\linewidth]{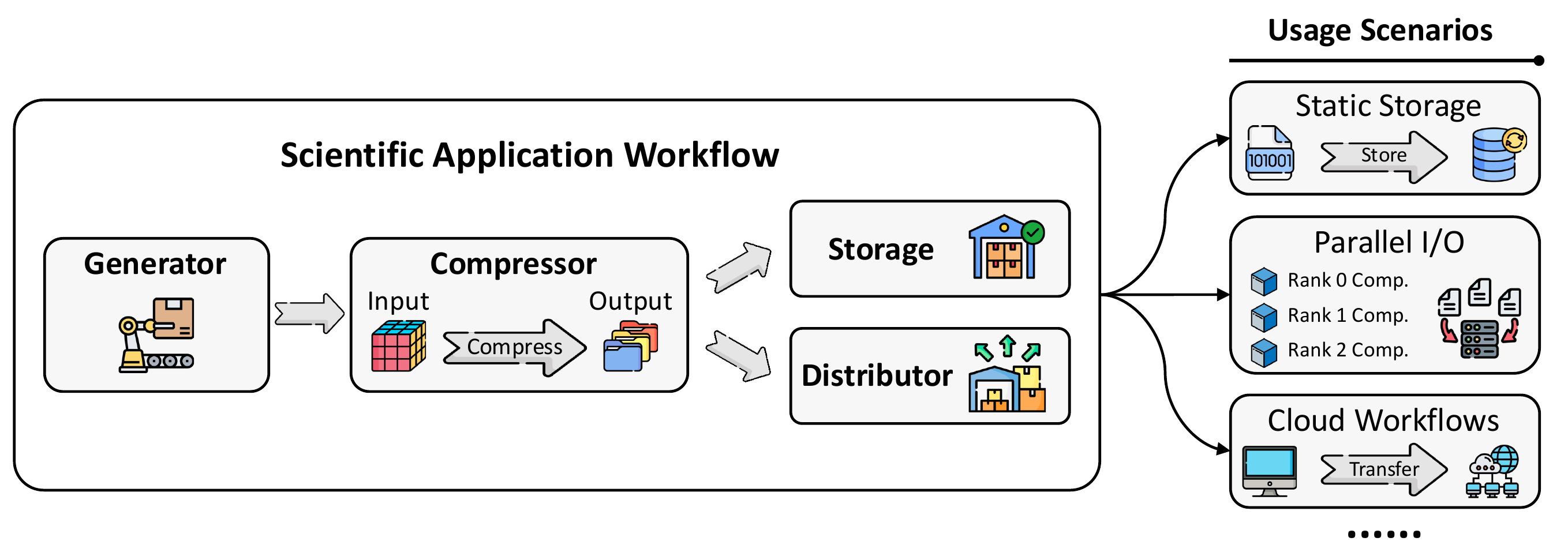}
    \caption{The use cases of error-bounded lossy compression in scientific applications.}
    \label{fig:background}
\end{figure}

Scientific applications routinely generate vast volumes of data for post-hoc analysis, visualization, and checkpointing~\cite{kay2015community,xu2025kilometer,skamarock2019description}. 
As application scale continues to increase, data management can consume a substantial fraction of the overall execution time and a large portion of system storage capacity~\cite{9844293,10046061}.
To support data management, scientific error-bounded lossy compression that allows users to specify a pointwise error bound has been incorporated into multiple stages of data-management workflows~\cite{sc25qoi,cappello2019use,cappello2025lossy,10046061}. Specifically, with a user-specified error bound $\Delta$, given input data $X$, compressed data $C$, and decompressed data $X^\prime$, error-bounded lossy compression guarantees that $\max|X-X^\prime|\leq \Delta$, and its compression ratio is the size of $X$ divided by the size of $C$ (so higher is better).
As Figure~\ref{fig:background} shows, when integrated into scientific applications, error-bounded lossy compression has several usages and brings benefits for scientific data management. Three representative use cases are:   
\begin{itemize}
    \item \textbf{Static Storage Reduction}: Storage reduction is a fundamental motivation for scientific data compression. To better benefit applications, compressors need to exhibit a high compression ratio.
    \item \textbf{Parallel I/O Acceleration}:  Scientific data compression may accelerate the data I/O of large-scale applications by performing parallelized data (de-)compression before accessing file systems. In this task, compared to the high local I/O bandwidth, compression throughput is the main bottleneck for I/O time optimization.
    \item \textbf{Cloud-Transfer Acceleration}: Scientific datasets can be lossy compressed before being transferred across distributed sites to accelerate the process. As network bandwidth is relatively lower than local I/O, a joint optimization of compression ratio and throughput is needed to sufficiently accelerate the end-to-end process.
\end{itemize}

In those workflows, a high compression ratio reduces the amount of data transferred and stored, whereas sufficiently high throughput is required to prevent compression from becoming part of the critical I/O path~\cite{10046061,10272494,11007768}. This requirement motivates broadly deployable compressors that can provide strong data reduction while scaling efficiently across threads, processes, and compute nodes~\cite{11007768,2025survey}.
    
\subsection{Problem Formulation and Analysis}
\label{sec:bg:formulation}
The three use cases above impose two coupled requirements on scientific lossy compression: effective data reduction and sufficiently high throughput. We next quantify these requirements and formulate the resulting research target.

\textbf{Research Target.}
Our research goal is to propose an error-bounded lossy compressor that jointly maximizes compression ratio and throughput to maintain storage efficiency and end-to-end speed. 
Moreover, its compression ratio and throughput should match the data-processing requirement of modern applications and systems.
We formulate this objective as:
\begin{equation}
\label{eq:research_target}
\begin{aligned}
\underset{\mathcal{A}}{\operatorname{maximize}}
\quad & T_c(\mathcal{A};X,\Delta,p),R(\mathcal{A};X,\Delta,p),p\in\mathcal{P},\\
\text{subject to}
\quad & R(\mathcal{A};X,\Delta,p) \geq R_{\mathrm{req}}(X,\Delta,p), p\in\mathcal{P},\\
& T_c(\mathcal{A};X,\Delta,p) \geq T_{c,\mathrm{req}}(X,\Delta,p), p\in\mathcal{P},\\ 
& \left\lVert X-\hat{X}_{\mathcal{A}}\right\rVert_{\infty}
       \leq \Delta .
\end{aligned}
\end{equation}
Here, $\mathcal{A}$ denotes the compressor, $X$ is the original dataset, and $\Delta$ is the user-specified error bound. $R(\mathcal{A};X,\Delta,p)$ and
$T_c(\mathcal{A};X,\Delta,p)$ denote the achieved compression ratio and compression throughput under execution parallelism $p$, respectively.
$R_{\mathrm{req}}(X,\Delta,p)$ and $T_{c,\mathrm{req}}(X,\Delta,p)$ denote the corresponding minimum compression ratio and throughput requirements imposed by the target application or system. Finally, $\mathcal{P}$ denotes the parallelism of execution.



\textbf{Compression-Ratio Requirements.}
A higher compression ratio directly translates to a reduced storage footprint. 
For an original data size $S$ and compression ratio $R$, the compressed size is $S/R$. Different scientific applications require compression ratios ranging from tens to hundreds~\cite{cappello2025lossy,sz3}. For example, seismological applications require an over $20\times$ reduction in data footprint to control data volume~\cite {sc25qoi}, yielding $R_{\mathrm{req}}(X,\Delta,p)\geq 20$. Earth-observation satellite systems require at least a $50\times$ reduction to relieve bandwidth-limited downlink bottlenecks, with targets growing alongside satellite constellations~\cite{sun2025deepspace}, yielding $R_{\mathrm{req}}(X,\Delta,p)\geq 50$.


\textbf{Throughput Requirements.}
For online data I/O and transfers, the compression ratio and throughput jointly determine the system performance. Let $T_c$ denote the compression throughput and $B$ represent the effective I/O bandwidth. The write or transfer times without ($T_{\mathrm{raw}}$) and with ($T_{\mathrm{comp}}$) compression can be modeled as:
$
T_{\mathrm{raw}}=\frac{S}{B},\ 
T_{\mathrm{comp}}=\frac{S}{T_c}+\frac{S}{RB}.
$
Therefore, compression reduces the end-to-end execution time if and only if:
$
T_c>\frac{B}{1-1/R} = T_{c,\min}.
$
On the compression ratio magnitude required by applications, $T_{c,\min}$ will be very close to the I/O bandwidth $B$ itself. 
\emph{System-Level Performance Analysis.} Representative scientific workloads ideally require an aggregate parallel-throughput of $10$~\text{GB/s}~\cite{cappello2025lossy}, which instantiates $T_{c,\mathrm{req}}(X,\Delta,p)\ge10~\text{GB/s}$ in Equation~\ref{eq:research_target}. Moreover, the IO500 submission for the leadership-class Aurora supercomputer reports an aggregate bandwidth of $934$~\text{GiB/s} utilizing $2{,}080$ client processes across ten nodes~\cite{io500AuroraSC23}. Normalizing this bandwidth by the number of client processes yields an average baseline per-process bandwidth of 
$B_{\mathrm{base}} \approx 482~\text{MB/s}$, so here it requires $T_{c,\mathrm{req}}(X,\Delta,1)\ge500~\text{MB/s}$ per core.

\section{Overview of \thiswork{}}
\label{sec:overview}

\begin{figure}[t]
    \centering
    \includegraphics[width=.9\linewidth]{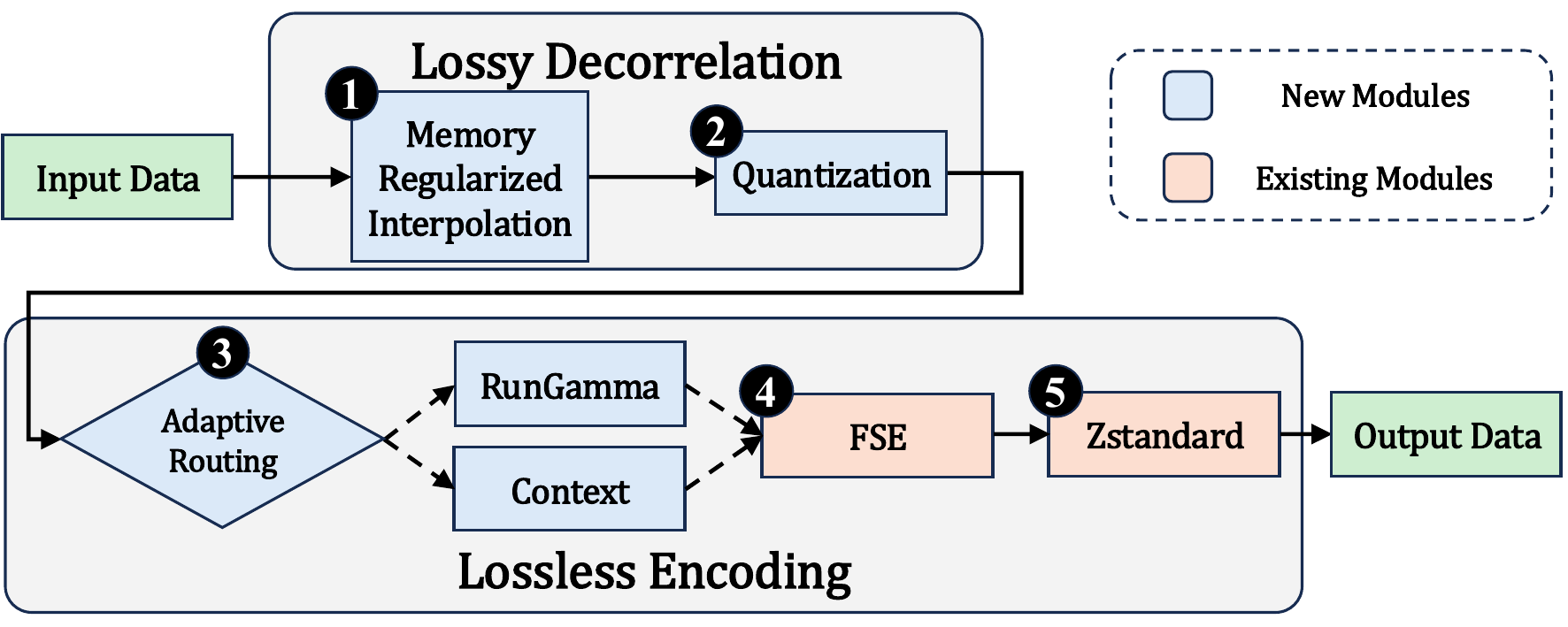}
    \caption{Overview of \thiswork{}.}
    \label{fig:overview}
\end{figure}

Our proposed \thiswork{} is a CPU-based scientific error-bounded lossy compressor, which concurrently achieves high compression ratio and ultra-fast throughput, ensuring universal availability across diverse platforms and applications. As depicted in Figure~\ref{fig:overview}, \thiswork{} follows a classical two-stage pipeline, composed of a redesigned interpolation-based data decorrelation stage that converts the input data into quantized integer errors and a distribution-adaptive lossless coding stage to pack the quantized errors. Outperforming existing works, \thiswork{} centers on three core novel designs:

\begin{itemize}
    \item \textbf{Memory-regularized interpolation predictor.}
    Dynamic spline interpolation has been adopted by several scientific compressors~\cite{szinterp,liu2022dynamic,liu2024high}. Unlike existing work, \thiswork{} eliminates irregular memory accesses and exposes contiguous prediction--quantization kernels to AVX2 and SVE2 vectorization, unlocking the computational potential of interpolation on modern CPUs.

    \item \textbf{Distribution-adaptive lossless coding.} \thiswork{} features a distribution-aware design of its entropy-centric lossless encoding pipeline. It dynamically dispatches quantization code streams to adaptive lossless pipelines, including our proposed processing modules \textit{RunGamma} and \textit{Context} plus the FSE (Finite-State Entropy)~\cite{fse} and Zstd~\cite{zstandard}, maximizing compression ratios and throughput on various data inputs and configurations.
    
    \item \textbf{Application-oriented parallel optimization.}
    \thiswork{} performs parallelization optimization to fit real-world scientific applications. In thread-independent schemes, it turns multi-pass interpolation into a low-traffic streaming dataflow, boosting cache locality and cutting memory traffic. For thread-cooperative parallelization, \thiswork{} employs fine-grained thread scheduling to eliminate contention without degrading spatial correlation.
\end{itemize}

Overall, \thiswork{} leverages the novel software-level design and hardware microarchitecture to \textbf{break the traditional trade-off between compression ratio and throughput}. Those approaches allow \thiswork{} to deliver consistent benefits for diverse requirements in parallel scientific applications.


\section{Core Design of ~\thiswork{}}
\label{sec:design}
To achieve significant acceleration for high-ratio scientific lossy compression in real-world use cases, \thiswork{} employs a bottom-up design: build the hardware-friendly computing framework first, then scale it across parallel environments. 
\thiswork{} first reconstructs the interpolation prediction and lossless pipeline, then introduces parallelization-oriented optimization for scalable execution.

\subsection{Memory Regularized Interpolation}
\begin{figure}[t]
    \centering
    \includegraphics[width=\linewidth]{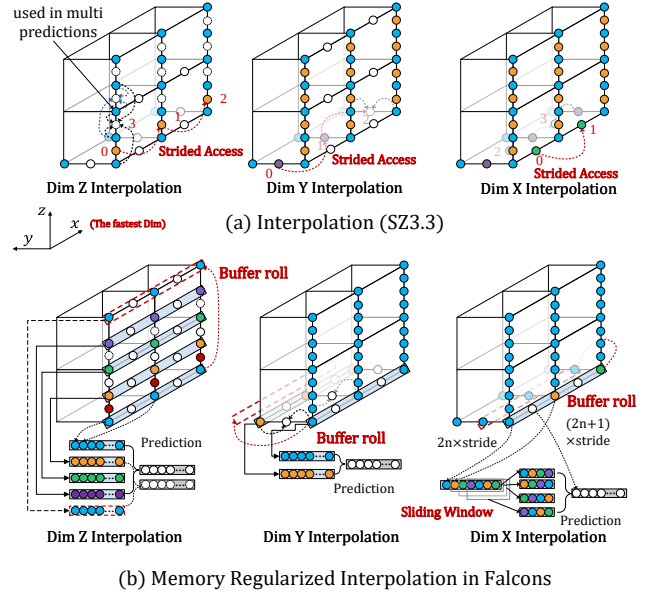}
    \caption{Comparison between interpolation-based data prediction in prior work and the memory-regularized, vectorized interpolation in \thiswork{}. Filled and hollow circles denote predictor and predictand values, respectively.}
    \label{fig:regular}
\end{figure}

Interpolation (dynamic spline interpolation) predicts scientific data hierarchically and progressively. Starting from a sparse grid of losslessly preserved anchor points, each interpolation level predicts previously unprocessed values from available adjacent data using a stride $s_\ell$, shown in Figure~\ref{fig:regular}(a). To ensure consistent compression and decompression, each prediction uses reconstructed neighboring data rather than the original data. The commonly used linear and cubic splines~\cite{szinterp,liu2022dynamic,liu2024high} are:
\[
p_i^{\mathrm{lin}}
=
\frac{\hat{x}_{i-s_\ell}+\hat{x}_{i+s_\ell}}{2},
p_i^{\mathrm{cubic}}
=
\frac{
-\hat{x}_{i-3s_\ell}
+9\hat{x}_{i-s_\ell}
+9\hat{x}_{i+s_\ell}
-\hat{x}_{i+3s_\ell}
}{16}.
\]
The prediction error is then quantized as
$q_i=\operatorname{round}((x_i-p_i)/(2\Delta))$,
and the reconstructed value is
$\hat{x}_i=p_i+2\Delta q_i$.

Although this hierarchical design effectively captures spatial correlations, its conventional implementation is poorly matched to modern architectures. As illustrated in Figure~\ref{fig:regular} (a), existing interpolation-based compressors exhibit two critical issues that will harm the computational throughput. \textbf{\textit{(1) Non-contiguous memory accesses:}} When moving across the predictand data (i.e., the data to be predicted), the accessed data index will increase by at least 2, which limits the use of hardware-level performance optimization techniques such as SIMD (Single Instruction, Multiple Data) and leads to unavoidable cache capacity misses.
\textbf{\textit{(2) Non-optimized reuse of predictor data:}} 
Predictor data values (i.e., the data used for prediction) are repeatedly accessed by subsequent interpolation stencils without an explicit mechanism for retaining and reusing them efficiently. 


\thiswork{} integrates an interpolation-based predictor that \textbf{reorganizes the interpolation computation to ensure regularized memory access}, while not modifying the prediction values to maintain accuracy.
As indicated in Figure~\ref{fig:regular} (b) as a 3-D example, the predictor data used for a 1-D subsequence along the fastest-varying dimension (namely, Dim X in Figure~\ref{fig:regular}) form a sparse structured sub-grid. Moreover, predictions within the same subsequence are independent. 
Therefore, \thiswork{} reorganizes those values into multiple aligned 1-D buffers residing in contiguous memory regions. Those buffers are along the fastest-varying dimension to minimize the build cost. 
\thiswork{} further maximally reuses buffered predictor data. For interpolation along the non-fastest-varying dimensions (e.g., Dim Z and Dim Y), it maintains multiple 1-D buffers—two for linear spline interpolation and four for cubic spline interpolation. 
After completing interpolation for the current predictand sequence and switching to a new one along the interpolation axis, the buffered window rolls forward, with one new buffer loaded and the others reused. 
For interpolation along the fastest-varying dimension (like Dim X in Figure~\ref{fig:regular} (b)), the predictand data and the predictor data interleave in the same 1-D sequence, so \thiswork{} performs a split of them, placing all predictor data in a contiguous 1-D buffer to enable efficient sliding-window-based interpolation.  

By transforming the strided memory accesses of interpolation into a contiguous pattern, \thiswork{} facilitates vectorized SIMD operations on modern CPUs to perform a fused prediction-quantization operation and leverages architecture-specific adaptations for AVX2/SVE2 to maximize efficiency.

\subsection{Quantization-Distribution Aware Lossless Encoding}

Existing interpolation-based error-bounded lossy compressors mostly apply a general-purpose Huffman encoder as their core entropy-coding module~\cite{Xin-bigdata18,szinterp,liu2022dynamic,liu2024cusz,liu2024high}, which exhibits two limitations in an application-aware, performance-oriented design: (1) the algorithmic nature of Huffman coding constrains its parallel scalability; (2) the generalized pipeline does not fully exploit the compressibility of quantized errors. After empirical study, algorithmic design, and performance optimization, \thiswork{} delivers a lossless coding pipeline composed of a core FSE (Finite-State Entropy) module, two newly designed preprocessors, and a lightweight distribution-aware online selection for the preprocessor.
\begin{table}[t]
\centering
\renewcommand{\arraystretch}{1}
\caption{Quantization-code statistics. $L_n$ denotes the average run length of code $n$. $H_0$ and $H_1$ denote the zero-order and first-order conditional entropy, respectively.}
\label{tab:quant-code-statistics}
\resizebox{.9\columnwidth}{!}{%
\begin{tabular}{lcccccc}
\toprule
\textbf{Field}
& \textbf{Error}
& \textbf{$L_0$}
& \textbf{$L_1$}
& \textbf{$L_{-1}$}
& \textbf{$H_0$}
& \textbf{$H_1$} \\
\midrule

temperature
& 1E-2 & \textbf{5,743.3} & 1.28 & 1.26 & 0.005 & \textbf{0.003} \\
temperature
& 1E-4 & \textbf{33.7} & 1.40 & 1.38 & 0.529 & \textbf{0.390} \\
\midrule

QV
& 1E-2 & \textbf{301.3} & 1.97 & 2.03 & 0.084 & \textbf{0.048} \\
QV
& 1E-4 & \textbf{6.8} & 1.33 & 1.27 & 2.517 & \textbf{2.038} \\
\midrule


TC48
& 1E-2 & \textbf{514.2} & 1.21 & 1.48 & 0.044 & \textbf{0.029} \\
TC48
& 1E-4 & \textbf{5.8} & 1.75 & 1.78 & 2.429 & \textbf{2.079} \\

\bottomrule
\end{tabular}%
}
\end{table}
\subsubsection{FSE as the Core Entropy-Coding Module}
As an entropy-coding algorithm, FSE~\cite{fse} is a table-based implementation of Asymmetric Numeral Systems (ANS)~\cite{duda2013asymmetric} that encodes symbols through transitions of a finite-state machine. It provides near-entropy compression with low per-symbol overhead, making it more attractive for high-throughput coding than Huffman. 
Prior work~\cite{adtfse} has demonstrated its potential for encoding quantized errors, but its broader adoption remains limited by inconsistent compression ratios and throughput across datasets. This variability arises because quantization code distributions change substantially with data characteristics and error bounds, ranging from low-entropy streams dominated by long symbol runs to high-entropy streams spread across broad symbol ranges.


\thiswork{} unlocks the full potential of FSE coding with the following intra-module and inter-module optimizations: 
(1) \thiswork{} partitions the quantization code sequence into multiple independent FSE streams and interleaves their state transitions to mitigate the loop-carried dependency between FSE state transitions that serializes the processing of a single coding stream. This design exposes instruction-level parallelism and improves the throughput.
(2) Adapting to the quantized error patterns, \thiswork{} leverages two different data preprocessing modules before the FSE process, ensuring the input data is friendly and efficient for FSE coding. 

\begin{table}[t]
\centering
\caption{Time costs and stage-wise compression ratio progression of different pipelines (data: Hurricane-TC, EB is error bound).}
\label{tab:lossless_comparison}
\setlength{\tabcolsep}{1.8pt}
\renewcommand{\arraystretch}{1}
\resizebox{\columnwidth}{!}{%
\begin{tabular}{@{}lcccccc|c@{}}
\toprule
\textbf{EB} & \textbf{Pipeline} & \shortstack{\textbf{Run}\\\textbf{coding}} & & \shortstack{\textbf{Entropy}\\\textbf{coding}} & & \textbf{Zstd} & \shortstack{\textbf{Time}\\\textbf{(ms)}} \\
\midrule

& Huffman + Zstd~\cite{Xin-bigdata18,sz3} & -- & & $31.7\times$ & $\stackrel{+2908.4\%}{\longrightarrow}$ & $952.6\times$ & 238 \\
& ADT-FSE + Zstd~\cite{adtfse} & -- & & $527.2\times$ & $\stackrel{+78.8\%}{\longrightarrow}$ & $942.6\times$ & 115 \\
\textbf{1E-2} & RLE + FSE + Zstd & $159.7\times$ & $\stackrel{+328.5\%}{\longrightarrow}$ & $684.2\times$ & $\stackrel{+0.0\%}{\longrightarrow}$ & $684.1\times$ & 18 \\
\rowcolor{selectedrow}
& \textbf{\texttt{RunGamma} + FSE + Zstd} & $159.7\times$ & $\stackrel{+592.5\%}{\longrightarrow}$ & $1106.0\times$ & $\stackrel{+6.8\%}{\longrightarrow}$ & $\mathbf{1181.3\times}$ & \textbf{17} \\
& \texttt{Context} + FSE + Zstd & -- & & $475.6\times$ & $\stackrel{+90.7\%}{\longrightarrow}$ & $906.9\times$ & 99 \\

\midrule

& Huffman + Zstd~\cite{Xin-bigdata18,sz3} & -- & & $24.2\times$ & $\stackrel{+139.8\%}{\longrightarrow}$ & $58.0\times$ & 355 \\
& ADT-FSE + Zstd~\cite{adtfse} & -- & & $36.0\times$ & $\stackrel{+14.6\%}{\longrightarrow}$ & $41.3\times$ & 140 \\
\textbf{1E-3} & RLE + FSE + Zstd & $6.8\times$ & $\stackrel{+370.6\%}{\longrightarrow}$ & $31.8\times$ & $\stackrel{+0.0\%}{\longrightarrow}$ & $31.8\times$ & 116 \\
\rowcolor{selectedrow}
& \textbf{\texttt{RunGamma} + FSE + Zstd} & $6.8\times$ & $\stackrel{+711.8\%}{\longrightarrow}$ & $55.2\times$ & $\stackrel{+11.0\%}{\longrightarrow}$ & $\mathbf{61.3\times}$ & \textbf{85} \\
& \texttt{Context} + FSE + Zstd & -- & & $56.7\times$ & $\stackrel{+3.7\%}{\longrightarrow}$ & $58.9\times$ & 110 \\

\midrule

& Huffman + Zstd~\cite{Xin-bigdata18,sz3} & -- & & $9.9\times$ & $\stackrel{+25.6\%}{\longrightarrow}$ & $12.5\times$ & 408 \\
& ADT-FSE + Zstd~\cite{adtfse} & -- & & $10.0\times$ & $\stackrel{+7.3\%}{\longrightarrow}$ & $10.8\times$ & 199 \\
\textbf{1E-4} & RLE + FSE + Zstd & $1.6\times$ & $\stackrel{+341.1\%}{\longrightarrow}$ & $6.9\times$ & $\stackrel{+0.1\%}{\longrightarrow}$ & $6.9\times$ & 461 \\
& \texttt{RunGamma} + FSE + Zstd & $1.6\times$ & $\stackrel{+656.3\%}{\longrightarrow}$ & $12.1\times$ & $\stackrel{+3.8\%}{\longrightarrow}$ & $12.6\times$ & 288 \\
\rowcolor{selectedrow}
& \textbf{\texttt{Context} + FSE + Zstd} & -- & & $15.4\times$ & $\stackrel{+1.5\%}{\longrightarrow}$ & $\mathbf{15.7\times}$ & \textbf{126} \\

\bottomrule
\end{tabular}%
}
\end{table}

\subsubsection{Distribution-Aware Quantization Data Preprocessing}
Quantization data from interpolation-based decorrelation generally exhibit two distinct forms: frequency skew (symbol probability distribution) and spatial redundancy (contiguous symbol repetitions). As shown in Table~\ref{tab:quant-code-statistics}, low-entropy streams contain long runs and highly skewed distributions, whereas high-entropy streams exhibit shorter runs and flatter distributions.

For low-entropy quantized data, we propose the \texttt{RunGamma} preprocessor, which comprises mode-based RLE (Run-length encoding) and Elias-gamma decomposition. \texttt{RunGamma} first explicitly captures spatial mode (0 in quantization codes) runs via mode-based RLE and transforms multi-byte run lengths into a more compressible representation to efficiently capture the frequency skew of non-zero symbols and run-length counts. Specifically, for any non-negative run-length $l$, \texttt{RunGamma} applies Elias-gamma decomposition to transform multi-bytes $l$ into a single-byte \emph{class} $k$ and a \emph{mantissa} $m$: $l+1 = 2^k + m, k = \left\lfloor\log_2(l+1)\right\rfloor, 0 \le m < 2^k.$ The class index k fits within a single byte and retains a highly skewed distribution, making it suitable for FSE encoding. The mantissa m is approximately uniform and can be directly bit-packed without entropy coding.
This design is inspired by our finding that the zero-run lengths and quantization codes closely approximate a geometric distribution and are inherently suited for Elias-gamma decomposition~\cite{1055349,kuznetsov2009eliasgamma,saukas2021zx0}. 
For high-entropy codes with short zero-runs, \thiswork{} includes the \texttt{Context} preprocessor, which models the conditional probabilities of quantization data symbols to further capture this dependency and reduce the entropy of the encoded data streams.
Adjacent quantized errors in the interpolation-ordered stream are produced from spatially related data points under similar prediction conditions, so they present solid correlation. Table~\ref{tab:quant-code-statistics} shows that 
quantization codes retain strong first-order conditional spatial correlation, yielding a significantly lower first-order conditional entropy than zero-order entropy (by 14.4\%-42.9\%).

\texttt{Context} first partitions a long quantization code stream into independently modeled segments to accommodate statistical variations across different interpolation levels and directional passes. Within each segment, the preceding code is mapped to a magnitude context bucket, $c_i=g(|q_{i-1}|)$, and an independent FSE table is constructed for each bucket to model the conditional distribution $P(q_i\mid c_i=c)$. During encoding, \texttt{Context} derives $c_i$ from $q_{i-1}$ and encodes the current code $q_i$ using the corresponding FSE table. 
Each segment stores its own context tables and encoded stream, allowing the model to adapt to local distribution changes. 

\begin{figure}[t]
\centering
\begin{subfigure}[t]{0.49\linewidth}
\centering
\includegraphics[width=\linewidth]{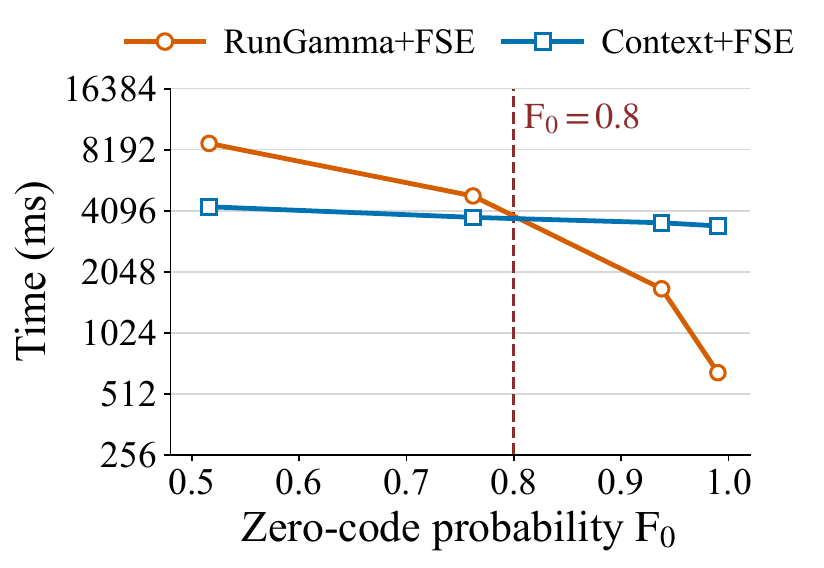}
\caption{}
\label{fig:adaptive_selection}
\end{subfigure}
\begin{subfigure}[t]{0.49\linewidth}
\centering
\definecolor{accuracyColor}{HTML}{0072B2}
\definecolor{lossColor}{HTML}{D55E00}
\resizebox{\linewidth}{!}{%
\begin{tikzpicture}[
    x=1cm,
    y=0.75cm,
    font=\sffamily\scriptsize,
    line cap=round,
    line join=round
]
\def\barWidth{1.02}
\def\xDataset{0.04}
\def\xAccuracy{1.78}
\def\xLoss{3.35}

\newcommand{\metriccell}[7]{%
    \pgfmathsetmacro{\filledWidth}
        {max(0,min(1,(#3-#4)/(#5-#4)))*\barWidth}
    \node[
        text=#6,
        font=\sffamily\scriptsize\bfseries
    ] at (#1,#2+0.07) {#7};
    \draw[gray!22,line width=1.5pt]
        (#1-\barWidth/2,#2-0.17) --
        (#1+\barWidth/2,#2-0.17);

    \draw[#6,line width=1.5pt]
        (#1-\barWidth/2,#2-0.17) --
        (#1-\barWidth/2+\filledWidth,#2-0.17);
    \fill[#6]
        (#1-\barWidth/2+\filledWidth,#2-0.17)
        circle[radius=1.35pt];
}
\node[
    anchor=west,
    font=\sffamily\scriptsize\bfseries
] at (\xDataset,4.12) {Dataset};
\node[
    font=\sffamily\scriptsize\bfseries,
    text=accuracyColor
] at (\xAccuracy,4.12) {Accuracy (\%)};

\node[
    font=\sffamily\scriptsize\bfseries,
    text=lossColor
] at (\xLoss,4.12) {Speed loss (\%)};

\draw[gray!35,line width=0.55pt]
    (0.02,3.87) -- (4.05,3.87);

\foreach \name/\yy in {
    NYX/3.48,
    Miranda/2.90,
    Hurricane/2.32,
    SCALE/1.74,
    CESM/1.16,
    JHTDB/0.58
}{
    \node[
        anchor=west,
        font=\sffamily\scriptsize
    ] at (\xDataset,\yy) {\name};
}

\foreach \yy in {3.19,2.61,2.03,1.45,0.87}{
    \draw[gray!14,line width=0.4pt]
        (0.02,\yy) -- (4.05,\yy);
}

\metriccell{\xAccuracy}{3.48}{100.00}{90}{100}
           {accuracyColor}{100.00}
\metriccell{\xLoss}{3.48}{0.000}{0}{0.6}
           {lossColor}{0.000}

\metriccell{\xAccuracy}{2.90}{97.14}{90}{100}
           {accuracyColor}{97.14}
\metriccell{\xLoss}{2.90}{0.086}{0}{0.6}
           {lossColor}{0.086}

\metriccell{\xAccuracy}{2.32}{95.00}{90}{100}
           {accuracyColor}{95.00}
\metriccell{\xLoss}{2.32}{0.542}{0}{0.6}
           {lossColor}{0.542}

\metriccell{\xAccuracy}{1.74}{98.33}{90}{100}
           {accuracyColor}{98.33}
\metriccell{\xLoss}{1.74}{0.179}{0}{0.6}
           {lossColor}{0.179}

\metriccell{\xAccuracy}{1.16}{97.97}{90}{100}
           {accuracyColor}{97.97}
\metriccell{\xLoss}{1.16}{0.079}{0}{0.6}
           {lossColor}{0.079}

\metriccell{\xAccuracy}{0.58}{100.00}{90}{100}
           {accuracyColor}{100.00}
\metriccell{\xLoss}{0.58}{0.000}{0}{0.6}
           {lossColor}{0.000}

\end{tikzpicture}%
}

\caption{}
\label{fig:selection_accuracy}
\end{subfigure}

\caption{Evaluation of the adaptive encoding selection mechanism.
(a) Variation of the zero-code probability and the execution times of
the two encoding methods. (b) Selection accuracy and the resulting
speed loss compared to the "oracle" choices.}
\label{fig:adaptive_selection_overview}
\end{figure}
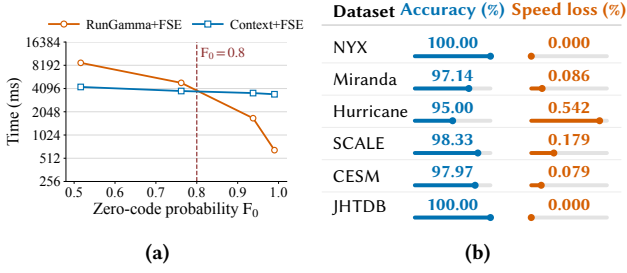

\subsubsection{Dynamic selection of preprocessors}
Table~\ref{tab:lossless_comparison} provides a quantitative analysis of different lossless pipelines with one representative scientific data input under different (error-bound) configurations. The key findings are twofold: (1) Compared to existing pipelines and naive preprocessing-based pipelines, the \texttt{RunGamma} and \texttt{Context} preprocessors substantially upgrade the efficiency of lossless coding for quantized errors, while bringing non-degraded compression ratios. (2) Under various use-case configurations, the performance advantages of \texttt{RunGamma} and \texttt{Context} can diverge.
To clearly distinguish the performance advantages of the two preprocessors, we thoroughly profiled them and confirmed the relationship between their performance and the top-1 (zero) symbol frequency. Figure~\ref{fig:adaptive_selection_overview}~(a) shows that, with a threshold of zero-symbol frequency $\approx$ 0.8, the performance rank of \texttt{RunGamma} and \texttt{Context} can be determined. To this end, in the online encoding of quantization data, \thiswork{} first profiles the zero-symbol frequency $F_0$ (in a sample-based routine with negligible overhead), and then dispatches the data to the better-fit preprocessor (\texttt{RunGamma} when $F_0 \geq 0.8$). Figure~\ref{fig:adaptive_selection_overview}~(b) demonstrates that this strategy achieves over 95\% selection accuracy, with less than a 1\% performance loss relative to oracle selection.



\subsection{Optimizing \thiswork{} in Parallel Scientific Applications}

When scientific data compression is integrated into applications and invoked in practical parallel environments, depending on the application tasks and parallelization schemes, new performance bottlenecks can arise, including but not limited to competition for memory bandwidth and cache capacity (in data-intensive weak-scaling cases) and compression-ratio degradation from reduced data access (in compute-intensive strong-scaling cases). Aware of these application-specific bottlenecks, \thiswork{} conducts systematic optimizations to fully enhance performance and scalability in diverse parallel scientific applications.



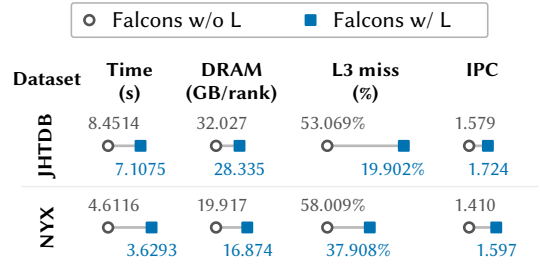
\begin{figure}[t]
\centering
\definecolor{baselineColor}{HTML}{555555}
\definecolor{optimizedColor}{HTML}{0072B2}
\begin{tikzpicture}[
    x=1cm,
    y=1cm,
    scale=1.08,
    transform shape,
    font=\sffamily\scriptsize,
    line cap=round,
    line join=round
]
\def\segmentScale{0.025}
\def\cacheSegmentScale{0.015}
\def\xDataset{0.43}
\def\xTime{1.18}
\pgfmathsetmacro{\xDram}{\xTime+\segmentScale*21.3+0.79}
\pgfmathsetmacro{\xCache}{\xDram+\segmentScale*15.3+0.98}
\pgfmathsetmacro{\xIpc}{\xCache+\cacheSegmentScale*62.5+0.80}

\pgfmathsetmacro{\xTimeHeader}{\xTime+\segmentScale*21.3/2}
\pgfmathsetmacro{\xDramHeader}{\xDram+\segmentScale*15.3/2}
\pgfmathsetmacro{\xCacheHeader}{\xCache+\cacheSegmentScale*62.5/2}
\pgfmathsetmacro{\xIpcHeader}{\xIpc+\segmentScale*13.3/2}

\newcommand{\metriccellmpiscaled}[8]{%
    \pgfmathsetmacro{\xBaseline}{#1}
    \pgfmathsetmacro{\xOptimized}{#1+#6*#3}
    \draw[gray!55, line width=1pt]
        (\xBaseline,#2) -- (\xOptimized,#2);
    \node[
        circle,
        draw=baselineColor,
        fill=white,
        line width=1pt,
        minimum size=3.4pt,
        inner sep=0pt
    ] at (\xBaseline,#2) {};
    \node[
        rectangle,
        draw=optimizedColor,
        fill=optimizedColor,
        line width=0.8pt,
        minimum size=3.4pt,
        inner sep=0pt
    ] at (\xOptimized,#2) {};
    \node[
        text=baselineColor,
        anchor=south,
        xshift=#7,
        yshift=2.2pt
    ] at (\xBaseline,#2) {#4};
    \node[
        text=optimizedColor,
        anchor=north,
        xshift=#8,
        yshift=-2.2pt
    ] at (\xOptimized,#2) {#5};
}

\newcommand{\metriccellmpi}[5]{%
    \metriccellmpiscaled{#1}{#2}{#3}{#4}{#5}
        {\segmentScale}{2pt}{0pt}%
}

\draw[gray!60, line width=0.6pt, rounded corners=1pt]
    (0.73,2.48) rectangle (5.83,2.90);
\node[
    circle,
    draw=baselineColor,
    fill=white,
    line width=1pt,
    minimum size=3.4pt,
    inner sep=0pt
] at (0.95,2.69) {};
\node[anchor=west,font=\sffamily\footnotesize]
    at (1.10,2.69) {\thiswork{} w/o \textsc{L}};
\node[
    rectangle,
    draw=optimizedColor,
    fill=optimizedColor,
    line width=0.8pt,
    minimum size=3.4pt,
    inner sep=0pt
] at (3.66,2.69) {};
\node[anchor=west,font=\sffamily\footnotesize]
    at (3.81,2.69) {\thiswork{} w/ \textsc{L}};

\node[anchor=north,font=\sffamily\scriptsize\bfseries]
    at (\xDataset,2.18) {Dataset};
\node[anchor=north,align=center,inner sep=0pt,
    font=\sffamily\scriptsize\bfseries]
    at (\xTimeHeader,2.18) {Time\\(s)};
\node[anchor=north,align=center,inner sep=0pt,
    font=\sffamily\scriptsize\bfseries]
    at (\xDramHeader,2.18) {DRAM\\(GB/rank)};
\node[anchor=north,align=center,inner sep=0pt,
    font=\sffamily\scriptsize\bfseries]
    at (\xCacheHeader,2.18) {L3 miss\\(\%)};
\node[anchor=north,align=center,inner sep=0pt,
    font=\sffamily\scriptsize\bfseries]
    at (\xIpcHeader,2.18) {IPC};

\node[rotate=90,font=\sffamily\scriptsize\bfseries]
    at (\xDataset,1.15) {JHTDB};
\node[rotate=90,font=\sffamily\scriptsize\bfseries]
    at (\xDataset,0.15) {NYX};
\draw[gray!15,line width=0.5pt]
    (0.12,0.65) -- (6.45,0.65);

\metriccellmpi{\xTime}{1.15}{15.9}{8.4514}{7.1075}
\metriccellmpi{\xDram}{1.15}{11.5}{32.027}{28.335}
\metriccellmpiscaled{\xCache}{1.15}{62.5}{53.069\%}{19.902\%}
    {\cacheSegmentScale}{2pt}{-4pt}
\metriccellmpi{\xIpc}{1.15}{9.2}{1.579}{1.724}

\metriccellmpi{\xTime}{0.15}{21.3}{4.6116}{3.6293}
\metriccellmpi{\xDram}{0.15}{15.3}{19.917}{16.874}
\metriccellmpiscaled{\xCache}{0.15}{34.7}{58.009\%}{37.908\%}
    {\cacheSegmentScale}{2pt}{-4pt}
\metriccellmpi{\xIpc}{0.15}{13.3}{1.410}{1.597}
\end{tikzpicture}
\caption{Compression performance and hardware metrics at 32 MPI ranks on NYX and JHTDB datasets. \textsc{L} denotes Low-Traffic Execution. Within each metric, segment length is proportional to the relative improvement from \thiswork{} w/o \textsc{L} to \thiswork{} w/ \textsc{L}.}
\label{fig:mpi}
\end{figure}

\subsubsection{Reducing Data Traffic for Weak-Scaling Parallel Workflows}
When each thread on the same chip or NUMA node processes a data snapshot independently for weak-scaling parallel tasks, CPU memory bandwidth and cache size become a critical bottleneck as parallelism increases.
Threads will compete for the shared L3 cache and memory channels, increasing cache evictions and memory-bandwidth pressure. 
To address this bottleneck, \thiswork{} fuses its interpolation operations along multiple dimensions into a low-traffic single-pass interpolation streaming dataflow.

For clarity, a three-dimensional case is provided as a representative example. At each interpolation level, \thiswork{} pipelines the directional interpolations to keep intermediate values cache-hot. Instead of completing a volume-wide $x$ pass before starting the $y$ pass, it processes each $y$ line as soon as its required $x$ lines become available, reducing the reuse distance from a full subdomain sweep to only a few neighboring lines. The resulting $x$--$y$ planes are then streamed into the $z$ interpolation according to their dependencies.
The same principle applies to two- to four-dimensional data by pipelining the available directional passes and retaining only the small set of slices required by their interpolation dependencies.
In addition, \thiswork{} requests transparent huge pages for each rank's quantization buffer to reduce dTLB misses under high rank density and eliminates reconstruction write-backs at the finest level when the written values will not be consumed by subsequent predictions.
As shown in Figure~\ref{fig:mpi}, on two test cases, these optimizations substantially reduce DRAM traffic and L3-cache miss rate while increasing IPC.

\subsubsection{Fine-grained Workload Partition for Strong Scaling Workflows}
Under strong scaling, an increasing number of threads cooperatively compress a fixed-size data snapshot.
Some existing compressors such as SZ3 and SPERR employ decorrelation algorithms that exploit global data redundancy. However, they straightforwardly partition input data into separate threads, which harms the performance of their decorrelation algorithms and lowers their compression-ratio relative to serial execution.
For example, given a data array $x_0,...,x_4,...,x_8$, partitioning may force $x_4$ to be predicted using only $x_0$, rather than both $x_0$ and $x_8$, reducing prediction accuracy.

\begin{figure}[t]
    \centering
    \includegraphics[width=\linewidth]{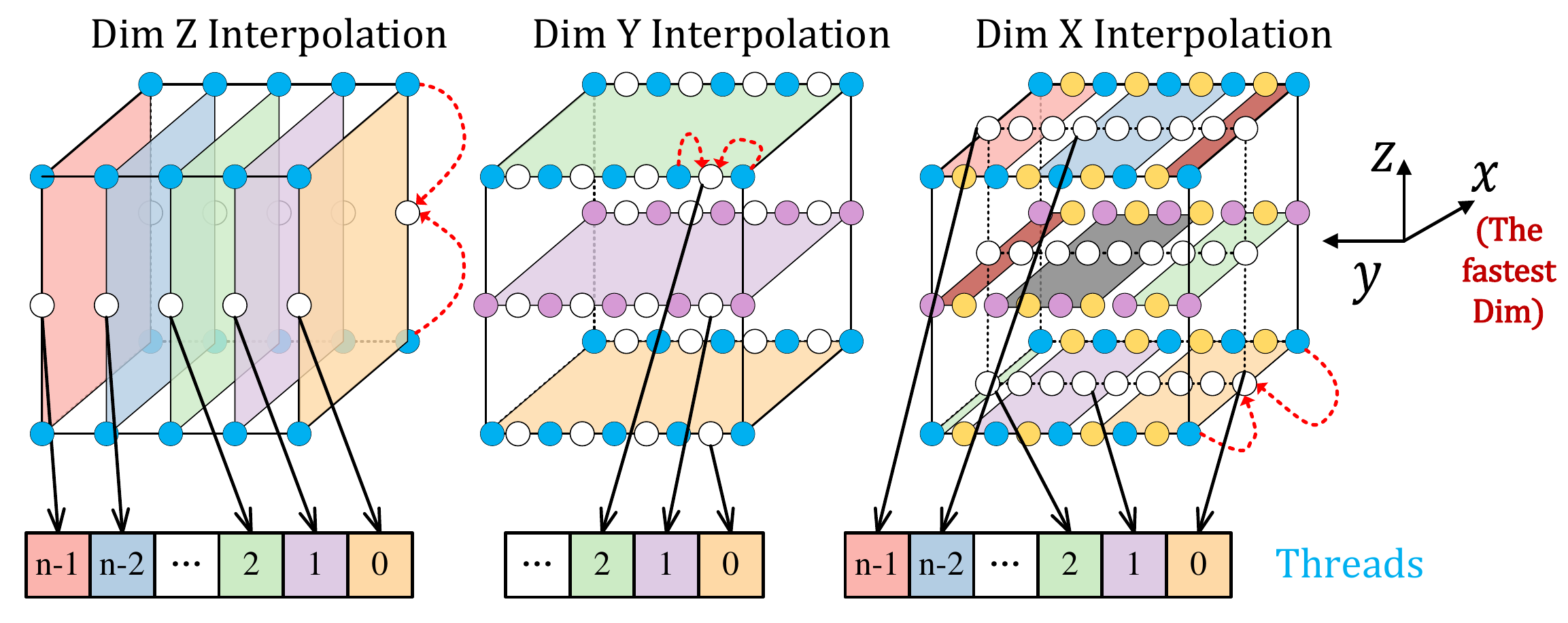}
    \caption{The fine-grained thread workload partition in \thiswork{}.}
    \label{fig:multithread}
\end{figure}
\begin{figure}[t]
    \centering
    \includegraphics[width=\linewidth]{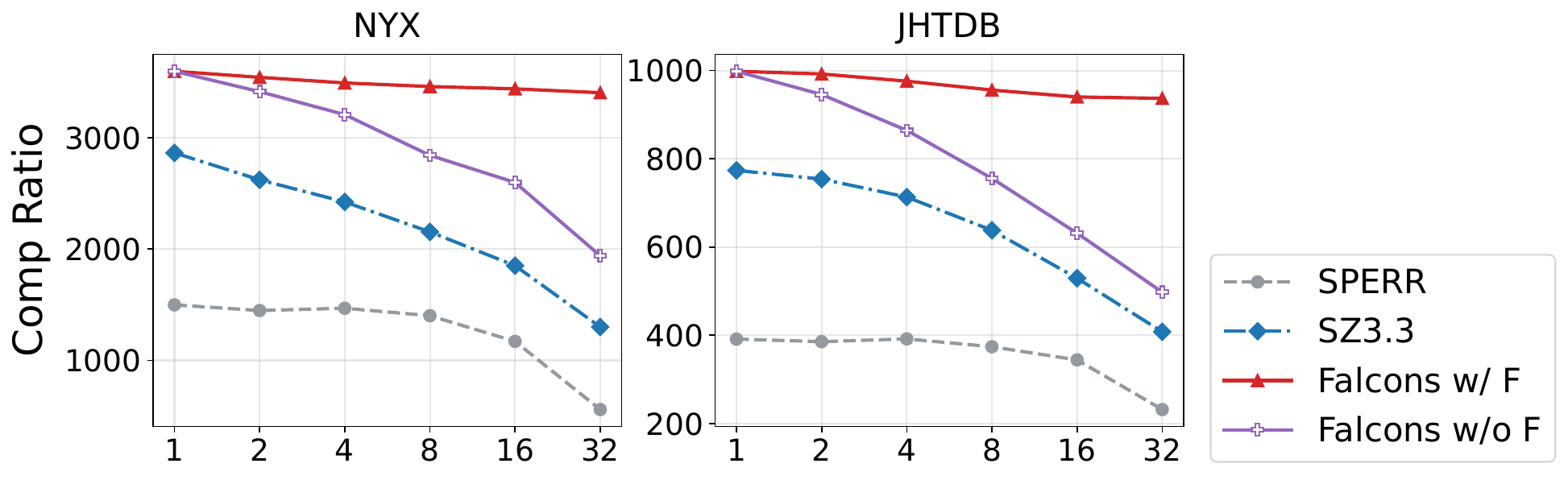}
    \caption{The compression ratios of compressors under strong-scaling parallelism at an error bound of 1E-2 on 2 representative datasets. \textsc{F} denotes Fine-grained workload partition. }
    \label{fig:ratio-omp}
\end{figure}

In contrast, \thiswork{} applies a fine-grained thread-level workload partition scheme, as demonstrated in Figure~\ref{fig:multithread} in a 3D example. To enable vectorization-aware regularized interpolation, the workload for each thread is contiguous along the current interpolation dimension and the fast-varying dimension. For example, for the first interpolation direction (Dim Z in Figure~\ref{fig:multithread}), each thread processes 2-D plane slices along Dim X (the fastest-varying dimension) and Dim Z (the current interpolation direction), and for the interpolation along the fastest-varying dimension X, each thread will process with a memory-contiguous data block along Dim X and Dim Y (not necessarily a 2-D slice). 
This workload partition scheme ensures a relative balance between threads, and keeps the same interpolation results as in the single-thread case. 
As illustrated in Figure~\ref{fig:ratio-omp}, several existing compressors and \thiswork{} without a fine-grained workload partition show a clear compression ratio drop in thread-cooperative compression, while \thiswork{} with fine-grained strong-scaling parallelism can preserve nearly identical compression ratio to the single-thread case (the slight drop is due to the parallelized lossless encoding).

\section{Evaluation}
\label{sec:eva}
In this section, we evaluate \thiswork{} thoroughly to validate its advantages in real-world scientific applications. 
Its performance is profiled on 6 different scientific domains and across cutting-edge supercomputers, from fundamental metrics to comprehensive use cases.

\subsection{Experimental Setup}
\textbf{\textit{Testbed.}} We conduct our experiments on two supercomputing platforms. \textit{Sugon-SD} supercomputer features two Intel Xeon Platinum 8358P (x86) 2.6 GHz processors with 64 physical cores and 512 GB of memory per computing node. \textit{LineShine} supercomputer~\cite{lineshine} features LX2 304C (ARM) CPUs at 1.55 GHz with 512 GB of memory per computing node. 
On Sugon-SD and LineShine, the compilers are GCC 9.3.0 and Lclang 1.0.0, respectively.

\textbf{\textit{Datasets.}} Datasets across six diverse scientific domains are adopted, which are widely used in previous studies~\cite{zhao2020sdrbench,liu2024high}. Detailed information on the datasets is provided in Table~\ref{tab:datasets}. The JHTDB dataset is collected from \cite{jhtdb}, and all other datasets are available at SDRBench~\cite{zhao2020sdrbench}.

\textbf{\textit{Baselines.}} We compare \thiswork{} against four state-of-the-art error-bounded scientific lossy compressors, spanning both ratio-oriented and throughput-oriented designs. As explained in Section~\ref{sec:intro}, GPU-based compressors are excluded because they target use scenarios different from those of CPU compressors.

\begin{itemize}
\item ZFP~\cite{lindstrom2014fixed}:  A representative transform-based scientific lossy compressor with a throughput-prioritized design.
\item PFPL~\cite{pfpl}: A fast scientific lossy compressor by efficiently encoding quantized input.
\item SPERR~\cite{sperr}: A high-ratio wavelet-transform-based scientific lossy compressor with superior compression quality.
\item SZ3.3~\cite{sz3}: The most recent version of the SZ compressor, which merges the old version of SZ3~\cite{szinterp} and HPEZ~\cite{liu2024high}.


\end{itemize}


\begin{table}[t]
\centering
\caption{Scientific datasets used in the evaluation.}
\label{tab:datasets}

\small
\setlength{\tabcolsep}{4pt}
\renewcommand{\arraystretch}{1.05}

\resizebox{1\linewidth}{!}{%
\begin{tabular}{@{}lll@{}}
\toprule
\textbf{Dataset} & \textbf{Application} & \textbf{Dimensions} \\
\midrule
CESM-ATM & Atmospheric simulation & $1800_y \times 3600_x$ \\

NYX & Cosmological simulation & $512_z \times 512_y \times 512_x$ \\

Miranda & Hydrodynamics simulation & $256_z \times 384_y \times 384_x$ \\

Hurricane ISABEL & Weather simulation & $100_z \times 500_y \times 500_x$ \\

SCALE-LETKF & Climate simulation & $98_z \times 1200_y \times 1200_x$ \\

JHTDB & Turbulence simulation & $512_z \times 512_y \times 512_x$ \\
\bottomrule
\end{tabular}%
}
\end{table}
\textbf{\textit{Configurations and Metrics.}} For the error-bound setting, we use the widely adopted value-range-based relative error bound (denoted by $\epsilon$), which is essentially equivalent to the absolute error bound (denoted by $\Delta$), following the relationship $\Delta = \epsilon \times \textit{value\_range}$. Each timing experiment is repeated five times, and the reported execution time is the arithmetic mean.
We perform the evaluation based on the following key metrics:
\begin{itemize}
    \item \textbf{Compression/Decompression Throughput}: Throughput shows the speed of compressors. It is the input data size divided by the compression/decompression time cost. 
    \item \textbf{Compression Ratio (CR) and bit rate}:  CR is defined as the original data size divided by the compressed data size. Bit rate is inversely proportional to CR and measures the average number of bits required to store each value. A higher compression ratio or a lower bit rate indicates more efficient storage.
    \item \textbf{Peak Signal-to-Noise Ratio (PSNR)}: PSNR~\cite{z-checker} quantifies the distortion between the input and decompressed data by value range and mean-square error, and a higher PSNR is better. 
    \item \textbf{Performance in real-world applications}: The data processing time cost in several scientific application use-case studies such as parallel data I/O, remote data transfer, and end-to-end integration analysis.
\end{itemize}

\begin{figure*}[t]
    \centering
    \begin{subfigure}{.96\textwidth}
        \centering
        \includegraphics[width=\textwidth]{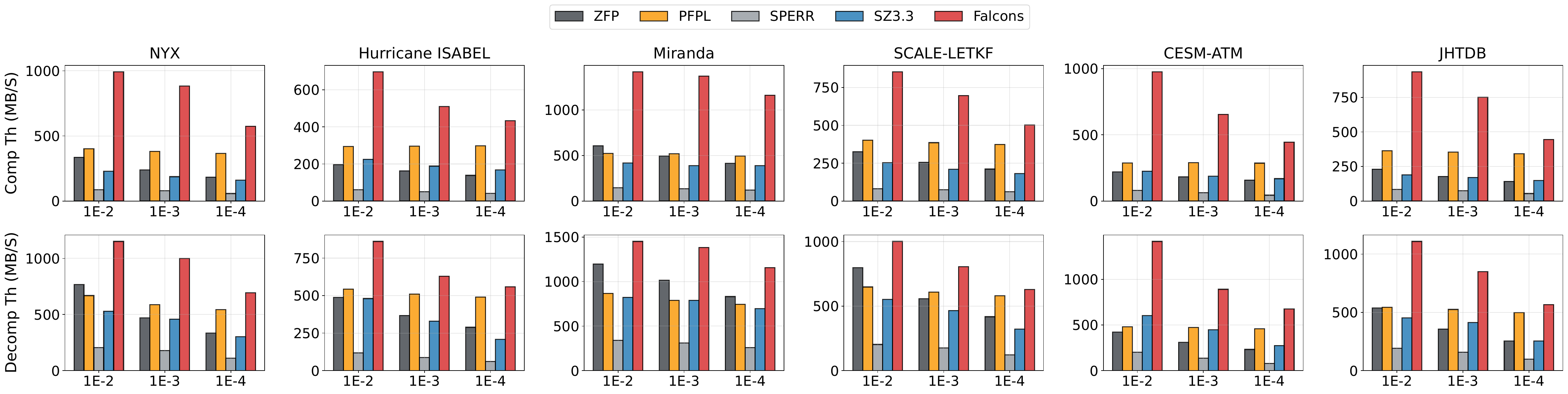}
        \caption{Intel Xeon Platinum 8358P (x86) on Sugon-SD}
    \end{subfigure}
    
    \begin{subfigure}{.96\textwidth}
        \centering
        \includegraphics[width=\textwidth]{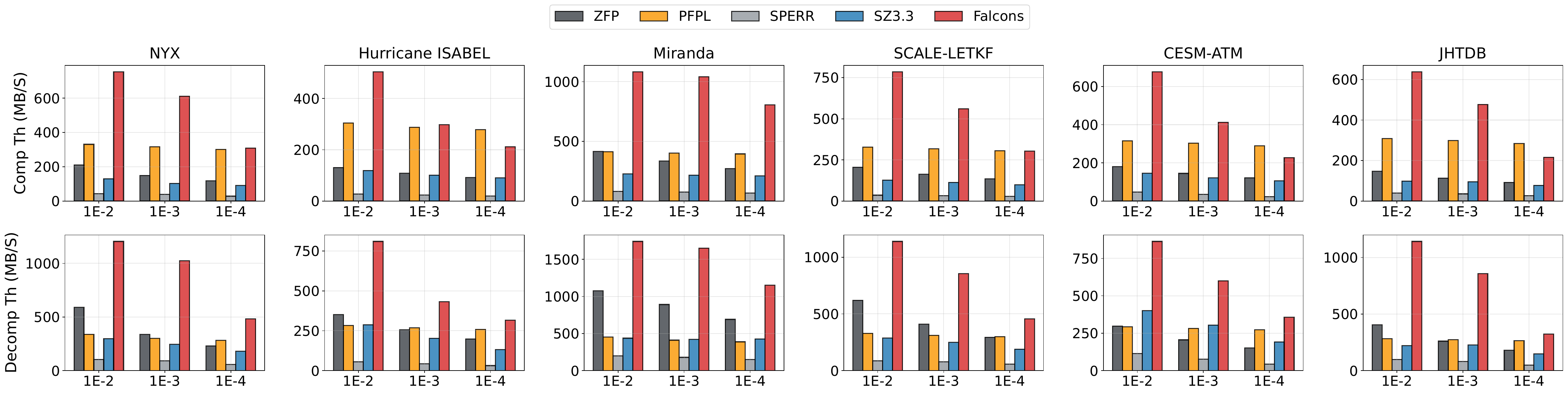}
         \caption{LX2 304C (ARM) on LineShine}
    \end{subfigure}
    \caption{The single-core compression and decompression throughput in MB/s on Intel and ARM CPUs under different error bounds.}
    \label{fig:serial-tp}
\end{figure*}

\subsection{Fundamental Performance Profiling}

\subsubsection{Ratio and Quality}
\begin{figure}[t]
    \centering
    \includegraphics[width=\linewidth]{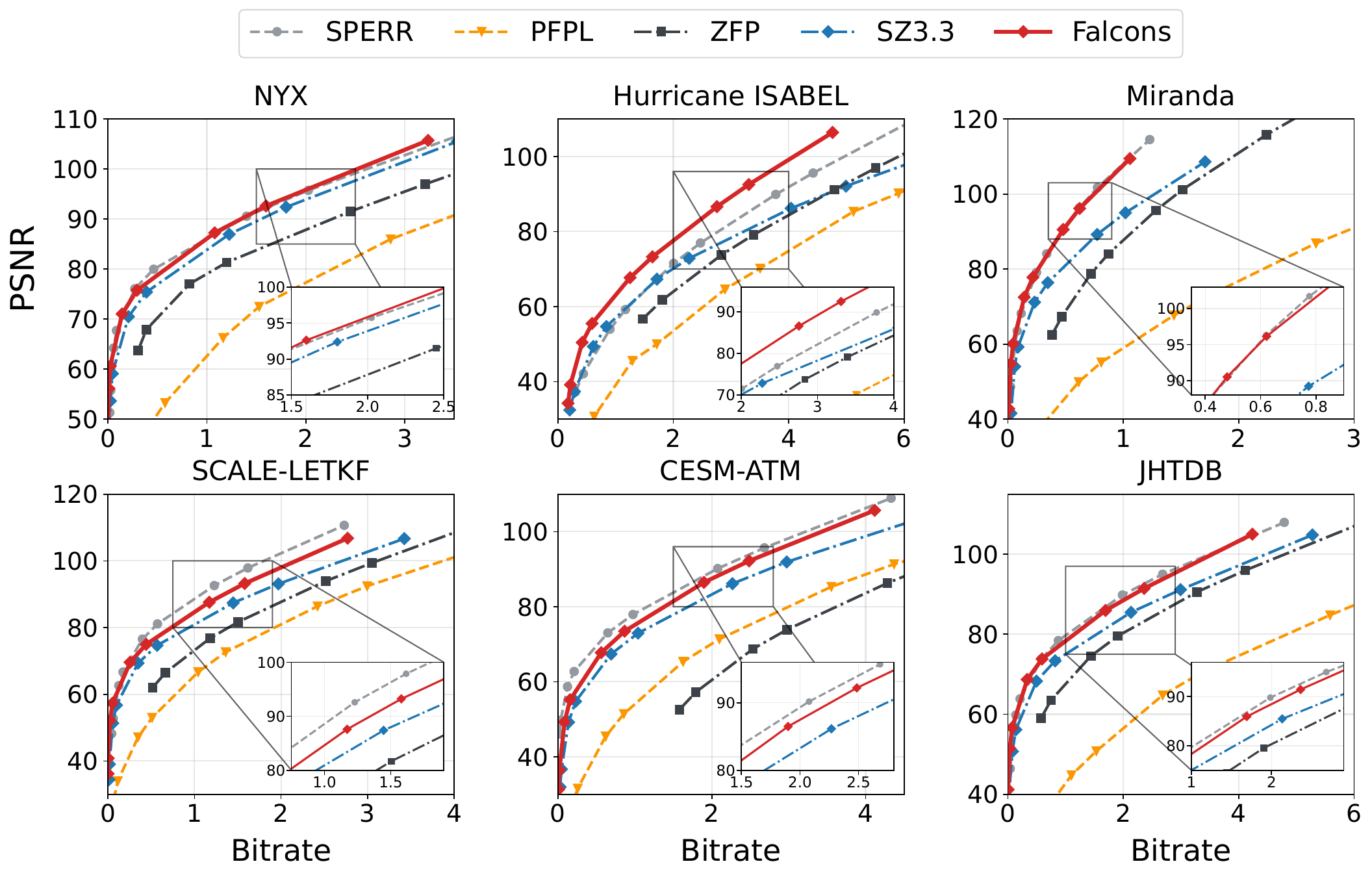}
    \caption{Rate Distortion (PSNR) Evaluation of \thiswork{} and other compressors under strong-scaling parallelism (32 threads). }
    \label{fig:rd}
\end{figure}

 At the beginning of the evaluation, we present the compression ratio and quality analysis of \thiswork{} together with other compressors. Table~\ref{tab:cr} presents the compression ratio of the evaluated scientific lossy compressors on all datasets under three representative error bounds. As a high-ratio scientific lossy compressor, \thiswork{} achieves up to a 34\%/155\% improvement over SZ3.3 and SPERR under the same error bound, and largely outperforms ZFP and PFPL in terms of compression ratio.  
 Figure~\ref{fig:rd} illustrates the rate-distortion (PSNR) of compressors in a 32-thread parallelism. It is also clear that  ~\thiswork{} achieves excellent rate-distortion, outperforming SZ3.3 and being comparable to SPERR (which is quite slower than ~\thiswork{}). These results justify the effective interpolation and encoding design of ~\thiswork{}, and in what follows, we can see that the high compression ratio of ~\thiswork{} is achieved without a throughput trade-off.

\subsubsection{Single-Core Throughput}
In Figure~\ref{fig:serial-tp}, the single-core throughput of the evaluated scientific compressors on two CPU architectures (x86 and ARM) is demonstrated.  In almost all test cases, \thiswork{} delivers the highest throughput, and in the few remaining ones, it achieves comparable speed to PFPL, which exhibits much lower compression ratios. Specifically, compared with high-ratio compressors, \thiswork{} achieves up to 547\%/418\% (to SZ3.3) and 2086\%/1213\% (to SPERR) compression/decompression throughput improvement. Moreover, \thiswork{} outperforms PFPL and ZFP by up to 255\%/306\% and 343\%/236\% in compression/decompression throughput. Featuring an interpolation-based design, \thiswork{} achieves $1.7\times$--$6.5\times$ speedups over SZ3.3, demonstrating the universal effectiveness of its memory-access-regularized interpolation and redesigned encoding pipeline across x86 and ARM CPU-based systems.

\subsubsection{Parallel Scalability}
\begin{figure}[t]
    \centering
    \begin{subfigure}{.9\linewidth}
        \centering
        \includegraphics[width=\linewidth]{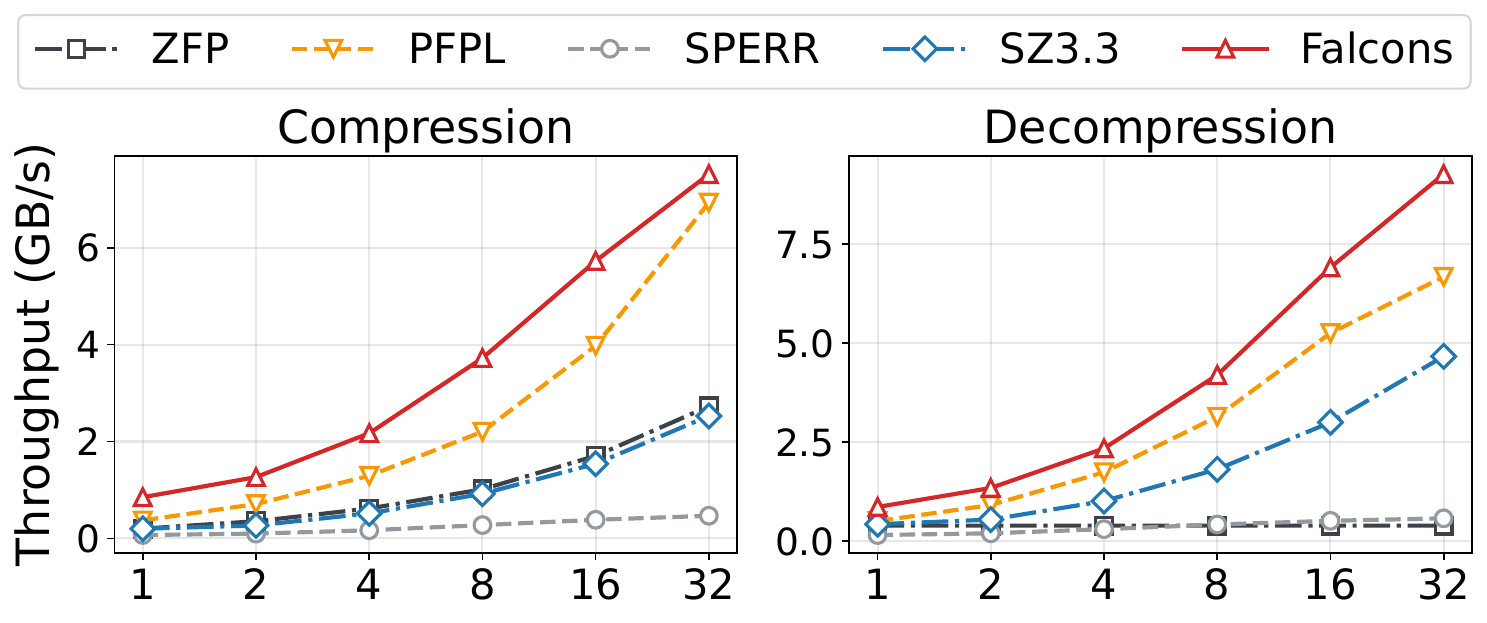}
        \caption{Throughput for strong-scaling parallelism}
        \label{fig:strong-scalability}
    \end{subfigure}

    \begin{subfigure}{.9\linewidth}
        \centering
        \includegraphics[width=\linewidth]{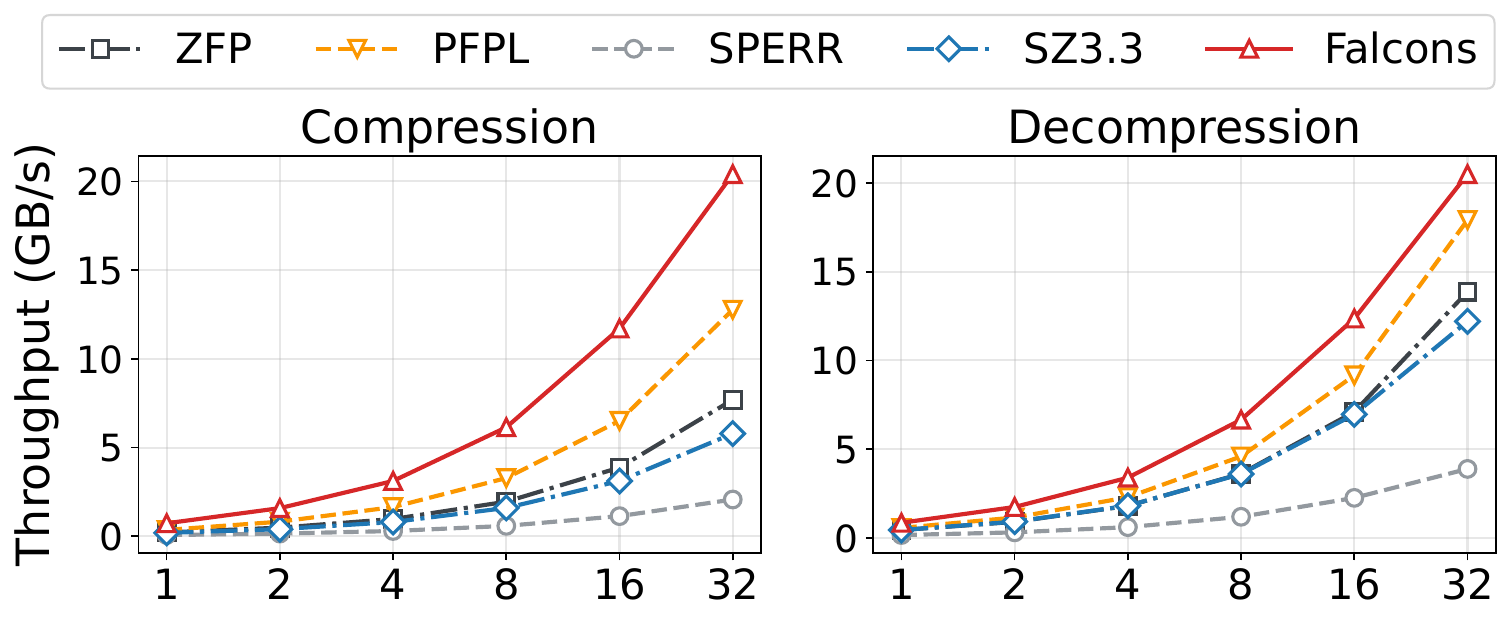}
         \caption{Throughput for weak-scaling parallelism}
         \label{fig:weak-scalability}
    \end{subfigure}
    \caption{The parallel compression and decompression throughput of compressors aggregated from six datasets.}
    \label{fig:scalability}
\end{figure}

Figure~\ref{fig:scalability} evaluates the scalability of \thiswork{} and
the baselines under two parallelization schemes.
Figure~\ref{fig:scalability}~(a) reports strong scaling on a fixed-size
input, while Figure~\ref{fig:scalability}~(b)
reports weak scaling using the increasing number of ranks with a fixed-size partition per rank. \thiswork{} consistently achieves the highest compression and
decompression throughput in both cases. Under strong scaling,
ratio-oriented fine-grained scheduling in \thiswork{} parallelizes independent interpolation lines while preserving the complete prediction context, increasing throughput with slightly degrading compression ratio, as shown in Figure~\ref{fig:ratio-omp}. Scaling becomes sublinear at high thread counts because of shrinking per-thread workloads and stage-level synchronization. Under weak scaling, the low-traffic streaming dataflow in \thiswork{} reduces repeated rank-local data traversals and shortens the reuse distance between dependent interpolation operations. This lowers the pressure placed by co-located ranks on the shared cache and memory channels, allowing aggregate compression and decompression throughput to grow steadily through a parallelism of 32.

\begin{table}[t]
\centering
\footnotesize
\setlength{\tabcolsep}{3pt}
\renewcommand{\arraystretch}{1}
\caption{Compression ratio of different compressors under single-core execution.  \textit{Improv} denotes improvement over the best baseline.}
\label{tab:cr}

\resizebox{0.9\columnwidth}{!}{%
\begin{tabular}{@{}c c
c c
>{\columncolor{gray!15}}c
>{\columncolor{gray!15}}c
>{\columncolor{gray!15}}c
c@{}}
\toprule
Dataset & $\epsilon$ & ZFP & PFPL
& SPERR & SZ3.3 & \thiswork{}
& Improv \\
\midrule

\multirow{3}{*}{NYX}
& 1E-2 & 38.8 & 83.9
& 1498.6 & 2863.6 & \textbf{\underline{3596.3}}
& $+25.6\%$ \\
& 1E-3 & 13.1 & 27.5
& 152.9 & 195.0 & \textbf{\underline{228.3}}
& $+17.1\%$ \\
& 1E-4 & 6.2 & 11.2
& 24.5 & 27.7 & \textbf{\underline{30.4}}
& $+9.7\%$ \\
\midrule

\multirow{3}{*}{Miranda}
& 1E-2 & 88.9 & 104.4
& 1434.9 & 2222.2 & \textbf{\underline{2606.0}}
& $+17.3\%$ \\
& 1E-3 & 49.9 & 44.4
& 400.4 & 392.3 & \textbf{\underline{482.2}}
& $+20.4\%$ \\
& 1E-4 & 28.6 & 24.0
& 138.1 & 104.9 & \textbf{\underline{140.5}}
& $+1.7\%$ \\
\midrule

\multirow{3}{*}{\makecell{Hurricane\\ISABEL}}
& 1E-2 & 11.3 & 24.8
& 32.5 & 76.1 & \textbf{\underline{78.2}}
& $+2.8\%$ \\
& 1E-3 & 6.7 & 11.0
& 15.2 & 23.2 & \textbf{\underline{26.1}}
& $+12.5\%$ \\
& 1E-4 & 4.3 & 6.2
& 8.4 & 9.6 & \textbf{\underline{11.8}}
& $+22.9\%$ \\
\midrule

\multirow{3}{*}{\makecell{SCALE-\\LETKF}}
& 1E-2 & 27.1 & 91.9
& 1137.1 & 1197.1 & \textbf{\underline{1383.4}}
& $+15.6\%$ \\
& 1E-3 & 12.7 & 30.7
& \textbf{\underline{158.6}} & 111.7 & 123.5
& $-22.1\%$ \\
& 1E-4 & 7.3 & 13.2
& \textbf{\underline{37.2}} & 24.4 & 28.6
& $-23.1\%$ \\
\midrule

\multirow{3}{*}{CESM-ATM}
& 1E-2 & 12.6 & 51.6
& 257.9 & 338.6 & \textbf{\underline{388.9}}
& $+14.9\%$ \\
& 1E-3 & 7.5 & 19.7
& 49.4 & 53.3 & \textbf{\underline{56.7}}
& $+6.4\%$ \\
& 1E-4 & 4.8 & 9.0
& 15.4 & 15.3 & \textbf{\underline{16.7}}
& $+8.4\%$ \\
\midrule

\multirow{3}{*}{JHTDB}
& 1E-2 & 22.3 & 29.0
& 391.4 & 773.8 & \textbf{\underline{998.4}}
& $+29.0\%$ \\
& 1E-3 & 9.8 & 11.9
& 69.5 & 82.5 & \textbf{\underline{97.8}}
& $+18.5\%$ \\
& 1E-4 & 5.0 & 5.7
& 18.1 & 16.4 & \textbf{\underline{20.2}}
& $+11.6\%$ \\
\bottomrule
\end{tabular}%
}
\end{table}


\subsection{Real-World Application Case Study}
After validating the fundamental performance metrics of \thiswork{}, we further conduct a series of case studies, covering the most representative usage of scientific lossy compression and demonstrating how it takes significant advantages over existing compressors in parallel scientific applications.

\subsubsection{Case Study I: Large-Scale Parallel Data I/O }
\begin{figure}[t]
    \centering
    \includegraphics[width=\linewidth]{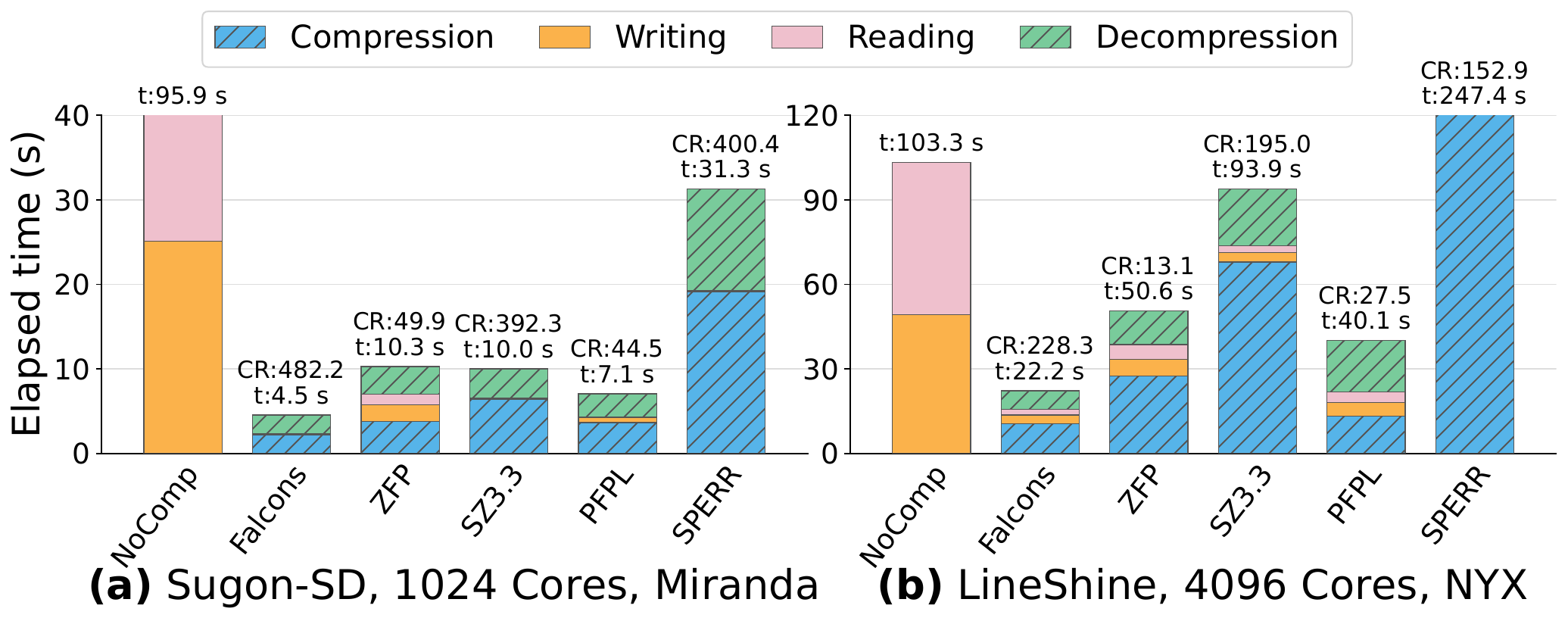}
    \caption{Large-scale parallel I/O performance with compression. }
    \label{fig:pio}
\end{figure}

In the first case, we directly integrate \thiswork{} into real-world scientific application workflows with large-scale parallelism. Specifically, \thiswork{} and other compressors are leveraged to accelerate the I/O process between large-scale parallel scientific applications and storage systems by performing error-bounded lossy compression/decompression before/after writing/reading the exascale scientific data stream to/from the storage.
Figure~\ref{fig:pio} presents the time costs and compression ratios of different compressors in this task. The performance evaluation is conducted with different scientific datasets on two supercomputers, Sugon-SD and LineShine, which exhibit distinct I/O bandwidths ($\approx$ 30 GB/s and $\approx$ 200 GB/s). Following the existing evaluation scheme~\cite{Xin-bigdata18,szinterp}, each parallel rank processes a copy of the dataset. Attributed to the extremely accelerated high-ratio design, in both cases, \thiswork{} achieves the minimized time cost for parallel I/O while reducing the storage cost to 0.21\% and 0.44\% of the original data size. On Sugon-SD with 1024 cores, \thiswork{} gains $\approx$ 37\% time reduction over the second-best PFPL, and for 4096 cores on LineShine, the corresponding reduction rate is $\approx$ 44.6\%. Those results explicitly demonstrate that, under varying hardware specifications and parallelism configurations, the performance advantage of \thiswork{} is consistent.

\subsubsection{Case Study II: Parallel Scientific Data Transfer Accelerated with Lossy Compression}

In the second use-case study, we target a scenario in which the data bandwidth is much more limited than local I/O, simulating the remote transfer of scientific data across different endpoints, which essentially supports distributed scientific analysis. Due to limited network bandwidth, scientific lossy compression can significantly accelerate the transfer process, as lossy-compressed data chunks can be delivered much faster than the original ones.
Table~\ref{tab:transfer_20x_total} profiles the end-to-end time (compression + transfer + decompression) of a compression-accelerated remote data transfer between Sugon supercomputing clusters at different locations. The network bandwidth is $\approx$ 1~GB/s, and threads on each node compress the data stream in a strong-scaling scheme. In all tested datasets, \thiswork{} achieves speedups over baselines, with performance gains of 51.5\% in maximum and 37.3\% across all datasets combined. Combining case studies I and II, we can confidently claim that \thiswork{} achieves SOTA performance in both computing-bounded and bandwidth-bounded data compression use cases, because its compression ratio and throughput are both leading among existing compressors.

\begin{table}[t]
\centering
\caption{End-to-end remote transfer time (in seconds) on Sugon using
20 nodes with 32 OpenMP threads each and an error bound of $10^{-3}$.
Transfer bandwidth is $\approx$ 1~GB/s. Gain denotes the percentage
time reduction of \thiswork{} over the fastest baseline.}
\label{tab:transfer_20x_total}
\renewcommand{\arraystretch}{1.0}
\setlength{\tabcolsep}{4pt}

\resizebox{\columnwidth}{!}{%
\begin{tabular}{@{}ccccccc@{}}
\toprule
Dataset & ZFP & PFPL & SPERR & SZ3.3 & Falcons & Speedup (\%) \\
\midrule
NYX & 128.3 & 33.8 & 60.9 & 25.3 & \textbf{12.3} & 51.5\% \\

Hurricane ISABEL & 89.4 & 30.5 & 67.3 & 30.1 & \textbf{19.7} & 34.5\% \\

Miranda & 39.0 & 16.3 & 27.5 & 11.6 & \textbf{7.2} & 37.9\% \\

SCALE-LETKF & 257.4 & 65.5 & 144.2 & 56.6 & \textbf{33.4} & 41.0\% \\

CESM-ATM & 130.5 & 29.5 & 376.8 & 51.3 & \textbf{27.3} & 7.5\% \\

JHTDB & 280.6 & 108.2 & 118.9 & 51.9 & \textbf{28.6} & 44.9\% \\
\bottomrule
\end{tabular}%
}
\end{table}
\subsubsection{Case Study III: End-to-End Analysis of \thiswork{} Integration into the WRF Workflow}

\begin{figure}[t]
    \centering
    \includegraphics[width=\linewidth]{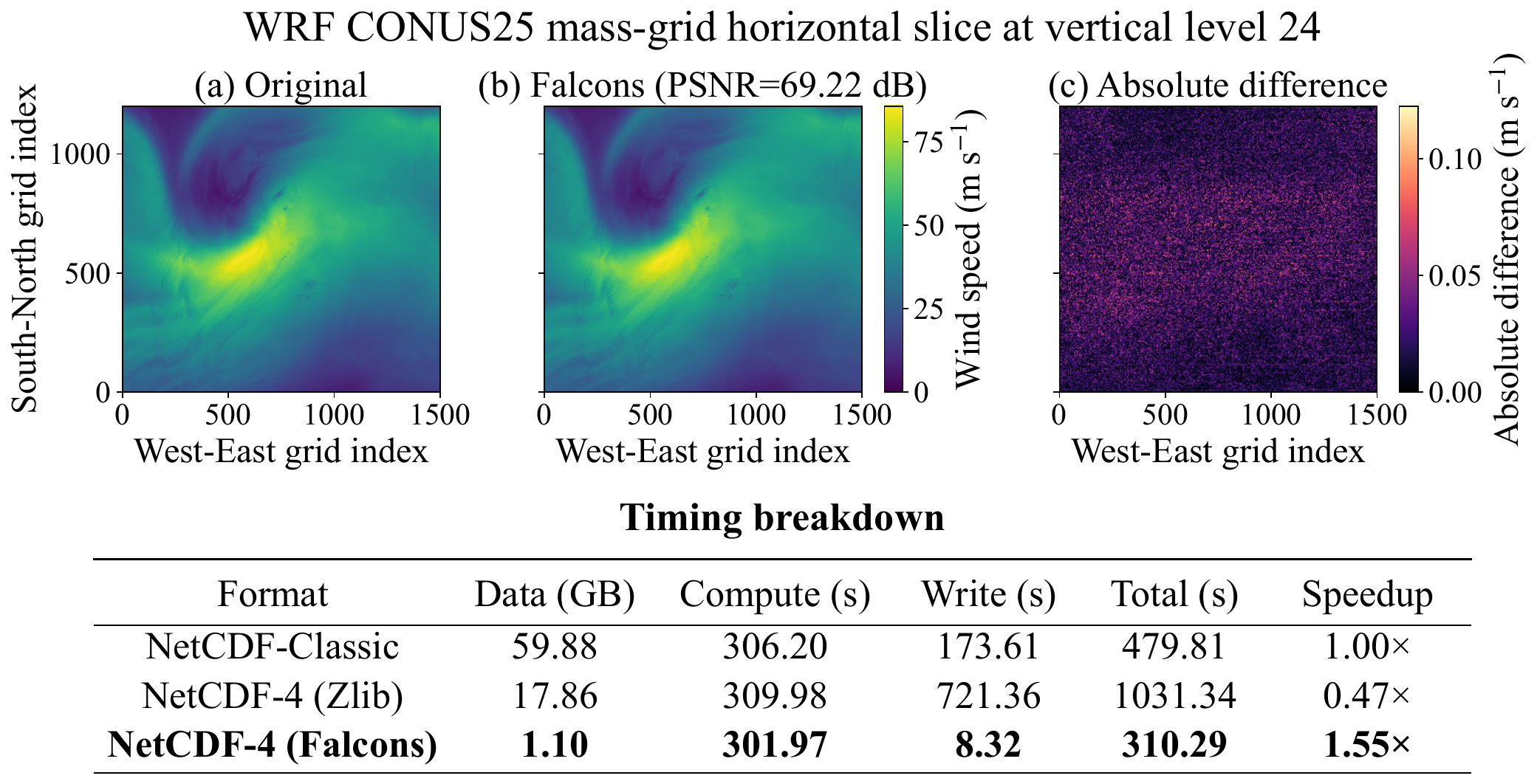}
    \caption{End-to-end analysis of \thiswork{} integration into the WRF. \thiswork{} compresses the data with the error bound of $10^{-3}$.}
    \label{fig:wrf}
\end{figure}

At the end of the case study and our evaluation, we demonstrate a practice of deeply integrating \thiswork{} into the WRF~\cite{skamarock2019description} (Weather Research \& Forecasting) application model. WRF is a widely used numerical weather prediction system for atmospheric research and forecasting, and it supports output in the NetCDF classic format~\cite{rew1990netcdf} (without data compression) and NetCDF-4~\cite{rew2006netcdf} format with Zlib~\cite{zlib} compression. We utilize the HDF5 filter to transparently integrate \thiswork{} into WRF’s NetCDF-4 I/O path, without modifying WRF’s numerical kernels.
In our evaluation, the CONUS 2.5-km benchmark~\cite{nersc_wrf_conus25} of WRF is executed on 2048 cores of the Sugon-SD supercomputer, and Figure~\ref{fig:wrf} profiles the end-to-end time costs with three different data filters. In the original WRF application, I/O time and data output size are two critical bottlenecks, and its native Zlib compression brings little help. By leveraging \thiswork{}, the two bottlenecks can be simultaneously resolved, reducing output data size to $\approx$ 1.8\% and accelerating end-to-end execution by 1.55$\times$.
Figure~\ref{fig:wrf} also visualizes a data slice of one physical field (wind speed) from the original output, \thiswork{} lossy decompressed output, together with the error map between them. Under the error bound of $10^{-3}$, \thiswork{} compressed data preserves satisfactory data quality (PSNR = 69.22) and a near-random error map. 

\section{Conclusion and Future Work}
\label{sec:con}
To overcome the performance bottleneck of scientific error-bounded lossy compression and advance its usability in real-world parallel scientific applications, we propose \thiswork{}, an ultra-fast high-ratio scientific error-bounded lossy compressor that breaks the trade-off between compression ratio and throughput. 
By introducing a memory-access regularized interpolation predictor and co-redesigning the de-correlation and lossless encoding modules, \thiswork{} delivers state-of-the-art compression ratio and throughput in the evaluation of 6 scientific domains. In real-world application case studies, compared to baselines, \thiswork{} gains up to 44.6\% and 51.5\% acceleration in large-scale parallel data I/O and remote transfer tasks. When integrated into the application as a data filter, it can reduce the end-to-end workflow time cost by 35\%.


To further enhance error-bounded lossy compression for real-world scientific application tasks in the future, we plan to: (1) exploit new-generation CPU hardware microarchitecture such as SME and AMX units for scientific data compression. (2) explore a more adaptive compression framework design that auto-selects the compression pipeline according to applications and hardware specifications.



\bibliographystyle{ACM-Reference-Format}
\bibliography{Reference}

@inproceedings{sun2025deepspace,
author = {Sun, Chuanhao and Zhang, Yu and Tao, Bill and Vasisht, Deepak and Marina, Mahesh},
title = {DeepSpace: Super Resolution Powered Efficient and Reliable Satellite Image Data Acquistion},
year = {2025},
isbn = {9798400715242},
publisher = {Association for Computing Machinery},
address = {New York, NY, USA},
url = {https://doi.org/10.1145/3718958.3750523},
doi = {10.1145/3718958.3750523},
booktitle = {Proceedings of the ACM SIGCOMM 2025 Conference},
pages = {311–328},
numpages = {18},
location = {S\~{a}o Francisco Convent, Coimbra, Portugal},
series = {SIGCOMM '25}
}

@ARTICLE{11007768,
  author={Liu, Yuanjian and Di, Sheng and Huang, Jiajun and Zhang, Zhaorui and Chard, Kyle and Foster, Ian},
  journal={IEEE Transactions on Parallel and Distributed Systems}, 
  title={Ocelot: An Interactive, Efficient Distributed Compression-As-a-Service Platform With Optimized Data Compression Techniques}, 
  year={2025},
  volume={36},
  number={12},
  pages={2497-2511},
  doi={10.1109/TPDS.2025.3568221}}

@INPROCEEDINGS{10272494,
  author={Liu, Yuanjian and Di, Sheng and Chard, Kyle and Foster, Ian and Cappello, Franck},
  booktitle={2023 IEEE 43rd International Conference on Distributed Computing Systems (ICDCS)}, 
  title={Optimizing Scientific Data Transfer on Globus with Error-Bounded Lossy Compression}, 
  year={2023},
  volume={},
  number={},
  pages={703-713},
  doi={10.1109/ICDCS57875.2023.00064}}

@ARTICLE{1055349,
  author={Elias, P.},
  journal={IEEE Transactions on Information Theory}, 
  title={Universal codeword sets and representations of the integers}, 
  year={1975},
  volume={21},
  number={2},
  pages={194-203},
  doi={10.1109/TIT.1975.1055349}}

@misc{kuznetsov2009eliasgamma,
  author       = {Kuznetsov, Anatoliy},
  title        = {Elias Gamma Encoding},
  year         = {2009},
  howpublished = {BitMagic Library Technical Note},
  url          = {https://bitmagic.io/dgap-gamma},
  note         = {Updated January 2020; accessed July 22, 2026}
}

@misc{saukas2021zx0,
  author       = {Saukas, Einar},
  title        = {{ZX0}: Data Compressor for 8-bit Computers and Low-end Platforms},
  year         = {2021},
  howpublished = {Computer software and format specification},
  url          = {https://github.com/einar-saukas/ZX0},
  note         = {GitHub repository, BSD 3-Clause License; accessed July 22, 2026}
}

@misc{cappello2025lossy,
      title={Lossy Compression of Scientific Data: Applications Constrains and Requirements}, 
      author={Franck Cappello and Allison Baker and Ebru Bozda and Martin Burtscher and Kyle Chard and Sheng Di and Paul Christopher O Grady and Peng Jiang and Shaomeng Li and Erik Lindahl and Peter Lindstrom and Magnus Lundborg and Kai Zhao and Xin Liang and Masaru Nagaso and Kento Sato and Amarjit Singh and Seung Woo Son and Dingwen Tao and Jiannan Tian and Robert Underwood and Kazutomo Yoshii and Danylo Lykov and Yuri Alexeev and Kyle Gerard Felker},
      year={2025},
      eprint={2503.20031},
      archivePrefix={arXiv},
      primaryClass={astro-ph.IM},
      url={https://arxiv.org/abs/2503.20031}, 
}

@ARTICLE{9844293,
  author={Liu, Yuanjian and Di, Sheng and Zhao, Kai and Jin, Sian and Wang, Cheng and Chard, Kyle and Tao, Dingwen and Foster, Ian and Cappello, Franck},
  journal={IEEE Transactions on Parallel and Distributed Systems}, 
  title={Optimizing Error-Bounded Lossy Compression for Scientific Data With Diverse Constraints}, 
  year={2022},
  volume={33},
  number={12},
  pages={4440-4457},
  doi={10.1109/TPDS.2022.3194695}}

@inproceedings{rew2006netcdf,
  author = {Rew, Russell K. and Hartnett, Ethan and Caron, J.},
  title = {NetCDF-4: Software Implementing an Enhanced Data Model for the Geosciences},
  booktitle = {22nd International Conference on Interactive Information Processing Systems for Meteorology, Oceanography, and Hydrology},
  volume = {6},
  pages = {395},
  year = {2006}
}

@article{rew1990netcdf,
  title={NetCDF: an interface for scientific data access},
  author={Rew, Russ and Davis, Glenn},
  journal={IEEE computer graphics and applications},
  volume={10},
  number={4},
  pages={76--82},
  year={1990},
  publisher={IEEE}
}

@INPROCEEDINGS{10046061,
  author={Jin, Sian and Tao, Dingwen and Tang, Houjun and Di, Sheng and Byna, Suren and Lukic, Zarija and Cappello, Franck},
  booktitle={SC22: International Conference for High Performance Computing, Networking, Storage and Analysis}, 
  title={Accelerating Parallel Write via Deeply Integrating Predictive Lossy Compression with HDF5}, 
  year={2022},
  volume={},
  number={},
  pages={1-15},
  doi={10.1109/SC41404.2022.00066}}

@inproceedings{adtfse,
author = {Lu, Tao and Zhong, Yu and Sun, Zibin and Chen, Xiang and Zhou, You and Wu, Fei and Yang, Ying and Huang, Yunxin and Yang, Yafei},
title = {ADT-FSE: A New Encoder for SZ},
year = {2023},
isbn = {9798400701092},
publisher = {Association for Computing Machinery},
address = {New York, NY, USA},
url = {https://doi.org/10.1145/3581784.3607044},
doi = {10.1145/3581784.3607044},
booktitle = {Proceedings of the International Conference for High Performance Computing, Networking, Storage and Analysis},
articleno = {45},
numpages = {13},
location = {Denver, CO, USA},
series = {SC '23}
}

@misc{io500AuroraSC23,
  author       = {{IO500 Foundation}},
  title        = {{Aurora}: SC23 IO500 Submission},
  year         = {2023},
  howpublished = {IO500 Submission No. 696},
  url          = {https://io500.org/submissions/view/696}
}

@misc{alcf2024aurora,
  author       = {{Argonne Leadership Computing Facility}},
  title        = {Overview of Aurora Exascale Architecture},
  year         = {2024},
  howpublished = {Argonne National Laboratory},
  url          = {https://www.alcf.anl.gov/sites/default/files/2024-11/Overview-of-Aurora-Oct-2024.pdf}
}

@misc{elcapitan,
  author       = {{Lawrence Livermore National Laboratory}},
  title        = {Lawrence Livermore National Laboratory's El Capitan Verified as World's Fastest Supercomputer},
  year         = {2024},
  month        = nov,
  howpublished = {LLNL News},
  url          = {https://www.llnl.gov/article/52061/lawrence-livermore-national-laboratorys-el-capitan-verified-worlds-fastest-supercomputer},
  note         = {Accessed: 2026-07-25}
}

@misc{fugaku,
  author       = {{RIKEN Center for Computational Science}},
  title        = {About Fugaku},
  year         = {2020},
  howpublished = {\url{https://www.r-ccs.riken.jp/en/fugaku/about/}},
  note         = {RIKEN Center for Computational Science website, accessed July 11, 2026}
}

@misc{lineshine,
  author       = {{TOP500}},
  title        = {{LineShine} Debuts at No.~1 as the {TOP500} Enters a New Global Exascale Era},
  year         = {2026},
  month        = jun,
  howpublished = {\url{https://top500.org/news/lineshine-debuts-no-1-top500-enters-new-global-exascale-era/}},
  note         = {Accessed: 2026-07-11}
}

@misc{zlib,
  author       = {Jean-loup Gailly and Mark Adler},
  title        = {zlib Compression Library},
  howpublished = {\url{https://zlib.net/}},
  year         = {2026},
  note         = {Accessed: 2026-08-01}
}

@misc{nersc_wrf_conus25,
  author       = {{National Energy Research Scientific Computing Center}},
  title        = {{WRF Benchmark on NERSC Systems: CONUS 2.5-km}},
  year         = {2023},
  howpublished = {\url{https://docs.nersc.gov/applications/wrf/wrf_benchmark/}},
  note         = {Accessed: 2026-08-01}
}

@techreport{skamarock2019description,
  title       = {A Description of the Advanced Research {WRF} Version 4},
  author      = {Skamarock, William C. and Klemp, Joseph B. and
                 Dudhia, Jimy and Gill, David O. and Liu, Zhiquan and
                 Berner, Judith and Wang, Wei and Powers, Jordan G. and
                 Duda, Michael G. and Barker, Dale M. and
                 Huang, Xiang-Yu},
  institution = {National Center for Atmospheric Research},
  number      = {NCAR/TN-556+STR},
  year        = {2019},
  pages       = {145},
  doi         = {10.5065/1dfh-6p97},
  url         = {https://doi.org/10.5065/1dfh-6p97}
}

@ARTICLE{lindstrom2014fixed,
  author={Lindstrom, Peter},
  journal={IEEE Transactions on Visualization and Computer Graphics}, 
  title={Fixed-Rate Compressed Floating-Point Arrays}, 
  year={2014},
  volume={20},
  number={12},
  pages={2674-2683},
  doi={10.1109/TVCG.2014.2346458},
  url={https://doi.org/10.1109/TVCG.2014.2346458}}

@inproceedings{di2016fast,
  title={Fast error-bounded lossy HPC data compression with SZ},
  author={Di, Sheng and Cappello, Franck},
  booktitle={2016 ieee international parallel and distributed processing symposium (ipdps)},
  pages={730--739},
  year={2016},
  organization={IEEE},
  url={https://doi.org/10.1109/IPDPS.2016.11}
}

@inproceedings{tao2017significantly,
  title={Significantly improving lossy compression for scientific data sets based on multidimensional prediction and error-controlled quantization},
  author={Tao, Dingwen and Di, Sheng and Chen, Zizhong and Cappello, Franck},
  booktitle={2017 IEEE International Parallel and Distributed Processing Symposium (IPDPS)},
  pages={1129--1139},
  year={2017},
  organization={IEEE},
  url={https://doi.org/10.1109/IPDPS.2017.115}
}

@inproceedings{liu2022dynamic,
  title={Dynamic quality metric oriented error bounded lossy compression for scientific datasets},
  author={Liu, Jinyang and Di, Sheng and Zhao, Kai and Liang, Xin and Chen, Zizhong and Cappello, Franck},
  booktitle={SC22: International Conference for High Performance Computing, Networking, Storage and Analysis},
  pages={1--15},
  year={2022},
  organization={IEEE},
  url={https://doi.org/10.1109/SC41404.2022.00067}
}

@article{liu2024high,
  title={High-performance effective scientific error-bounded lossy compression with auto-tuned multi-component interpolation},
  author={Liu, Jinyang and Di, Sheng and Zhao, Kai and Liang, Xin and Jin, Sian and Jian, Zizhe and Huang, Jiajun and Wu, Shixun and Chen, Zizhong and Cappello, Franck},
  journal={Proceedings of the ACM on Management of Data},
  volume={2},
  number={1},
  pages={1--27},
  year={2024},
  publisher={ACM New York, NY, USA},
  url={https://doi.org/10.1145/3639259}
}

@inproceedings{lu2023zfp,
  title={Zfp-x: Efficient embedded coding for accelerating lossy floating point compression},
  author={Lu, Bing and Li, Yida and Wang, Junqi and Luo, Huizhang and Li, Kenli},
  booktitle={2023 IEEE International Parallel and Distributed Processing Symposium (IPDPS)},
  pages={1041--1050},
  year={2023},
  organization={IEEE},
  url={https://doi.org/10.1109/IPDPS54959.2023.00107}
}

@article{zfp2025,
author = {Peter Lindstrom and Jeffrey Hittinger and James Diffenderfer and Alyson Fox and Daniel Osei-Kuffuor and Jeffrey Banks},
title ={ZFP: A compressed array representation for numerical computations},

journal = {The International Journal of High Performance Computing Applications},
volume = {39},
number = {1},
pages = {104-122},
year = {2025},
doi = {10.1177/10943420241284023},

URL = { 
        https://doi.org/10.1177/10943420241284023
},
eprint = { 
        https://doi.org/10.1177/10943420241284023
}
}

@inproceedings{lu2026prism,
  title={PRISM: An Efficient GPU-Based Lossy Compression Framework for Progressive Data Retrieval with Multi-Level Interpolation},
  author={Lu, Bing and Liu, Zedong and Zhao, Hairui and Luo, Dejun and Huang, Wenjing and Gu, Yida and Liu, Jinyang and Tan, Guangming and Tao, Dingwen},
  booktitle={Proceedings of the 31st ACM SIGPLAN Annual Symposium on Principles and Practice of Parallel Programming},
  pages={164--176},
  year={2026},
  url={https://doi.org/10.1145/3774934.3786438}
}

@article{chen2025computational,
  title={Computational Burst Buffers: Accelerating HPC I/O via In-Storage Compression Offloading},
  author={Chen, Xiang and Lu, Bing and Long, Haoquan and Luo, Huizhang and Ma, Yili and Tan, Guangming and Tao, Dingwen and Wu, Fei and Lu, Tao},
  journal={IEEE Transactions on Parallel and Distributed Systems},
  volume={37},
  number={2},
  pages={518--532},
  year={2025},
  publisher={IEEE}
}

@inproceedings{hdf5,
author = {Folk, Mike and Heber, Gerd and Koziol, Quincey and Pourmal, Elena and Robinson, Dana},
title = {An overview of the HDF5 technology suite and its applications},
year = {2011},
isbn = {9781450306140},
publisher = {Association for Computing Machinery},
address = {New York, NY, USA},
url = {https://doi.org/10.1145/1966895.1966900},
doi = {10.1145/1966895.1966900},
booktitle = {Proceedings of the EDBT/ICDT 2011 Workshop on Array Databases},
pages = {36–47},
numpages = {12},
location = {Uppsala, Sweden},
series = {AD '11}
}

@inproceedings{grosset2020foresight,
  title={Foresight: Analysis that matters for data reduction},
  author={Grosset, Pascal and Biwer, Christopher M and Pulido, Jesus and Mohan, Arvind T and Biswas, Ayan and Patchett, John and Turton, Terece L and Rogers, David H and Livescu, Daniel and Ahrens, James},
  booktitle={SC20: International Conference for High Performance Computing, Networking, Storage and Analysis},
  pages={1--15},
  year={2020},
  organization={IEEE},
  url={https://doi.org/10.1109/SC41405.2020.00087}
}

@article{cappello2019use,
  title={Use cases of lossy compression for floating-point data in scientific data sets},
  author={Cappello, Franck and Di, Sheng and Li, Sihuan and Liang, Xin and Gok, Ali Murat and Tao, Dingwen and Yoon, Chun Hong and Wu, Xin-Chuan and Alexeev, Yuri and Chong, Frederic T},
  journal={The International Journal of High Performance Computing Applications},
  volume={33},
  number={6},
  pages={1201--1220},
  year={2019},
  publisher={Sage Publications Sage UK: London, England},
  url={https://doi.org/10.1177/1094342019853336}
}

@inproceedings{tian2020cusz,
author = {Tian, Jiannan and Di, Sheng and Zhao, Kai and Rivera, Cody and Fulp, Megan Hickman and Underwood, Robert and Jin, Sian and Liang, Xin and Calhoun, Jon and Tao, Dingwen and Cappello, Franck},
title = {cuSZ: An Efficient GPU-Based Error-Bounded Lossy Compression Framework for Scientific Data},
year = {2020},
isbn = {9781450380751},
publisher = {Association for Computing Machinery},
address = {New York, NY, USA},
url = {https://doi.org/10.1145/3410463.3414624},
doi = {10.1145/3410463.3414624},
booktitle = {Proceedings of the ACM International Conference on Parallel Architectures and Compilation Techniques},
pages = {3–15},
numpages = {13},
location = {Virtual Event, GA, USA},
series = {PACT '20}
}

@article{MGARD,
  title={Multilevel techniques for compression and reduction of scientific data—the univariate case},
  author={Ainsworth, Mark and Tugluk, Ozan and Whitney, Ben and Klasky, Scott},
  journal={Computing and Visualization in Science},
  volume={19},
  number={5},
  pages={65--76},
  year={2018},
  publisher={Springer},
  url={https://doi.org/10.1007/s00791-018-00303-9}
}

@ARTICLE{liang2021mgard,
author={Liang, Xin and Whitney, Ben and Chen, Jieyang and Wan, Lipeng and Liu, Qing and Tao, Dingwen and Kress, James and Pugmire, David and Wolf, Matthew and Podhorszki, Norbert and Klasky, Scott},
journal={ IEEE Transactions on Computers },
title={{MGARD+: Optimizing Multilevel Methods for Error-Bounded Scientific Data Reduction}},
year={2022},
volume={71},
number={07},
ISSN={1557-9956},
pages={1522-1536},
doi={10.1109/TC.2021.3092201},
url = {https://doi.ieeecomputersociety.org/10.1109/TC.2021.3092201},
publisher={IEEE Computer Society},
address={Los Alamitos, CA, USA},
month=jul}

@inproceedings{zhao2020sdrbench,
  title={{SDRB}ench: Scientific data reduction benchmark for lossy compressors},
  author={Zhao, Kai and Di, Sheng and Lian, Xin and Li, Sihuan and Tao, Dingwen and Bessac, Julie and Chen, Zizhong and Cappello, Franck},
  booktitle={IEEE International Conference on Big Data},
  pages={2716--2724},
  year={2020},
  url={https://doi.org/10.1109/BigData50022.2020.9378449}
}

@inproceedings{huang2023cuszp,
  title={cuszp: An ultra-fast gpu error-bounded lossy compression framework with optimized end-to-end performance},
  author={Huang, Yafan and Di, Sheng and Yu, Xiaodong and Li, Guanpeng and Cappello, Franck},
  booktitle={Proceedings of the International Conference for High Performance Computing, Networking, Storage and Analysis},
  pages={1--13},
  year={2023},
  url={https://doi.org/10.1145/3581784.3607048}
}

@inproceedings{huang2024cuszp2,
  title={cuSZp2: A GPU lossy compressor with extreme throughput and optimized compression ratio},
  author={Huang, Yafan and Di, Sheng and Li, Guanpeng and Cappello, Franck},
  booktitle={SC24: International Conference for High Performance Computing, Networking, Storage and Analysis},
  pages={1--18},
  year={2024},
  organization={IEEE},
  url={https://doi.org/10.1109/SC41406.2024.00021}
}

@inproceedings{liu2024cusz,
author = {Liu, Jinyang and Tian, Jiannan and Wu, Shixun and Di, Sheng and Zhang, Boyuan and Underwood, Robert and Huang, Yafan and Huang, Jiajun and Zhao, Kai and Li, Guanpeng and Tao, Dingwen and Chen, Zizhong and Cappello, Franck},
title = {cuSZ-i: High-Ratio Scientific Lossy Compression on GPUs with Optimized Multi-Level Interpolation},
year = {2024},
isbn = {9798350352917},
publisher = {IEEE Press},
url = {https://doi.org/10.1109/SC41406.2024.00019},
doi = {10.1109/SC41406.2024.00019},
booktitle = {Proceedings of the International Conference for High Performance Computing, Networking, Storage, and Analysis},
articleno = {13},
numpages = {15},
location = {Atlanta, GA, USA},
series = {SC '24}
}

@article{kay2015community,
  title={The Community Earth System Model (CESM) Large Ensemble Project: A Community Resource for Studying Climate Change in the Presence of Internal Climate Variability},
  author={Kay, Jennifer E and Deser, Clara and Phillips, A and Mai, A and Hannay, Cecile and Strand, Gary and Arblaster, Julie Michelle and Bates, S. C. and Danabasoglu, Gokhan and Edwards, James and Holland, Marika and Kushner, Paul and Lamarque, Jean-Fran{\c{c}}ois and Lawrence, David and Lindsay, Keith and Middleton, Andrew and Munoz, Esmond and Neale, Richard and Oleson, Keith and Polvani, Lorenzo and Vertenstein, Mariana},
  journal={Bulletin of the American Meteorological Society},
  volume={96},
  number={8},
  pages={1333--1349},
  year={2015},
  doi={10.1175/BAMS-D-13-00255.1}
}

@inproceedings{xu2025kilometer,
author = {Xu, Kai and Yu, Maoxue and Chen, Yuhu and Gao, Jie and Wang, Shuang and Song, Jiaying and Duan, Xiaohui and Wei, Junlin and Yu, Jiangfeng and Liu, Hailong and Jiang, Jinrong and Zhang, Yi and Lin, Pengfei and Wang, Tianyi and Wang, Pengfei and Zheng, Weipeng and Xie, Jingwei and Zhang, Jiakang and Liu, Zilu and Jin, Xiaoyu and Wei, Jilin and Chang, Qixin and Lin, Qingxia and Zhou, Yanzhi and Liu, Weiguo and Xue, Wei and Li, Yiwen and Fu, Haohuan and Yu, Yue and Chi, Xuebin and Wu, Lixin},
title = {Kilometer-Scale AI-Powered and Performance-Portable Earth System Model (AP3ESM) to Achieve Year-Scale Simulation Speed on Heterogeneous Supercomputers},
year = {2025},
isbn = {9798400714665},
publisher = {Association for Computing Machinery},
address = {New York, NY, USA},
url = {https://doi.org/10.1145/3712285.3771788},
doi = {10.1145/3712285.3771788},
booktitle = {Proceedings of the International Conference for High Performance Computing, Networking, Storage and Analysis},
pages = {111–124},
numpages = {14},
location = {
},
series = {SC '25}
}

@article{cappello2025multifacets,
title = {Multifacets of lossy compression for scientific data in the Joint-Laboratory of Extreme Scale Computing},
journal = {Future Generation Computer Systems},
volume = {163},
pages = {107323},
year = {2025},
issn = {0167-739X},
doi = {https://doi.org/10.1016/j.future.2024.05.022},
url = {https://www.sciencedirect.com/science/article/pii/S0167739X24002553},
author = {Franck Cappello and Mario Acosta and Emmanuel Agullo and Hartwig Anzt and Jon Calhoun and Sheng Di and Luc Giraud and Thomas Grützmacher and Sian Jin and Kentaro Sano and Kento Sato and Amarjit Singh and Dingwen Tao and Jiannan Tian and Tomohiro Ueno and Robert Underwood and Frédéric Vivien and Xavier Yepes and Yoshii Kazutomo and Boyuan Zhang}
}

@INPROCEEDINGS{sperr,
  author={Li, Shaomeng and Lindstrom, Peter and Clyne, John},
  booktitle={2023 IEEE International Parallel and Distributed Processing Symposium (IPDPS)}, 
  title={Lossy Scientific Data Compression With SPERR}, 
  year={2023},
  volume={},
  number={},
  pages={1007-1017},
  doi={10.1109/IPDPS54959.2023.00104},
  url={https://doi.org/10.1109/IPDPS54959.2023.00104}}

@INPROCEEDINGS{pfpl,
  author={Fallin, Alex and Azami, Noushin and Di, Sheng and Cappello, Franck and Burtscher, Martin},
  booktitle={2025 IEEE International Parallel and Distributed Processing Symposium (IPDPS)}, 
  title={Fast and Effective Lossy Compression on GPUs and CPUs with Guaranteed Error Bounds}, 
  year={2025},
  volume={},
  number={},
  pages={874-887},
  doi={10.1109/IPDPS64566.2025.00083},
  url={https://doi.org/10.1109/IPDPS64566.2025.00083}}

@article{ballester2019tthresh,
  title     = {{TTHRESH}: Tensor compression for multidimensional visual data},
  author    = {Ballester-Ripoll, Rafael and Lindstrom, Peter and Pajarola, Renato},
  journal   = {IEEE transactions on visualization and computer graphics},
  volume    = {26},
  number    = {9},
  pages     = {2891--2903},
  year      = {2019},
  publisher = {IEEE},
  url = {https://doi.org/10.1109/TVCG.2019.2904063}
}

@article{duda2013asymmetric,
  title={Asymmetric numeral systems: entropy coding combining speed of huffman coding with compression rate of arithmetic coding},
  author={Duda, Jarek},
  journal={arXiv preprint arXiv:1311.2540},
  year={2013}
}

@misc{fse,
  author       = {Yann Collet},
  title        = {Finite State Entropy},
  year         = {2013},
  howpublished = {\url{https://github.com/Cyan4973/FiniteStateEntropy}},
  note         = {Software repository}
}

@article{2025survey,
author = {Di, Sheng and Liu, Jinyang and Zhao, Kai and Liang, Xin and Underwood, Robert and Zhang, Zhaorui and Shah, Milan and Huang, Yafan and Huang, Jiajun and Yu, Xiaodong and Ren, Congrong and Guo, Hanqi and Wilkins, Grant and Tao, Dingwen and Tian, Jiannan and Jin, Sian and Jian, Zizhe and Wang, Daoce and Rahman, Md Hasanur and Zhang, Boyuan and Song, Shihui and Calhoun, Jon and Li, Guanpeng and Yoshii, Kazutomo and Alharthi, Khalid and Cappello, Franck},
title = {A Survey on Error-Bounded Lossy Compression for Scientific Datasets},
year = {2025},
issue_date = {November 2025},
publisher = {Association for Computing Machinery},
address = {New York, NY, USA},
volume = {57},
number = {11},
issn = {0360-0300},
url = {https://doi.org/10.1145/3733104},
doi = {10.1145/3733104},
journal = {ACM Comput. Surv.},
month = jun,
articleno = {287},
numpages = {38}
}

@inproceedings{sc25qoi,
author = {Cappello, Franck and Underwood, Robert and Alexeev, Yuri and Baker, Alison and Bozda\u{g}, Ebru and Burtscher, Martin and Chard, Kyle and Di, Sheng and Felker, Kyle Gerard and O'Grady, Paul Christopher and Guo, Hanqi and Huang, Yafan and Jiang, Peng and Jin, Sian and Johansson, Petter and Li, Shaomeng and Liang, Xin and Lindahl, Erik and Lindstrom, Peter and Luki\'{c}, Zarija and Lundborg, Magnus and Lykov, Danylo and Nagaso, Masaru and Sato, Kento and Singh, Amarjit and Son, Seung Woo and Song, Shihui and Tang, William and Tao, Dingwen and Tian, Jiannan and Yoshii, Kazutomo and Zhao, Kai},
title = {What to Support When You’re Compressing: The State of Practice Gaps and Opportunities for Scientific Data Compression},
year = {2025},
isbn = {9798400714665},
publisher = {Association for Computing Machinery},
address = {New York, NY, USA},
url = {https://doi.org/10.1145/3712285.3759856},
doi = {10.1145/3712285.3759856},
booktitle = {Proceedings of the International Conference for High Performance Computing, Networking, Storage and Analysis},
pages = {1966–1979},
numpages = {14},
location = {
},
series = {SC '25}
}

@inproceedings{szinterp,
  author    = {Zhao, Kai and Di, Sheng and Dmitriev, Maxim and Tonellot, Thierry-Laurent D. and Chen, Zizhong and Cappello, Franck},
  booktitle = {2021 IEEE 37th International Conference on Data Engineering (ICDE)},
  title     = {Optimizing Error-Bounded Lossy Compression for Scientific Data by Dynamic Spline Interpolation},
  year      = {2021},
  pages     = {1643-1654},
  doi       = {10.1109/ICDE51399.2021.00145},
  url       = {https://doi.org/10.1109/ICDE51399.2021.00145}
}

@article{sz3,
  author   = {Liang, Xin and Zhao, Kai and Di, Sheng and Li, Sihuan and Underwood, Robert and Gok, Ali M. and Tian, Jiannan and Deng, Junjing and Calhoun, Jon C. and Tao, Dingwen and Chen, Zizhong and Cappello, Franck},
  journal  = {IEEE Transactions on Big Data},
  title    = {SZ3: A Modular Framework for Composing Prediction-Based Error-Bounded Lossy Compressors},
  year     = {2023},
  volume   = {9},
  number   = {2},
  pages    = {485-498},
  doi      = {https://doi.org/10.1109/TBDATA.2022.3201176},
  url = {https://doi.org/10.1109/TBDATA.2022.3201176}
}

@inproceedings{Xin-bigdata18,
  author       = {Xin Liang and Sheng Di and Dingwen Tao and Sihuan Li and Shaomeng Li and Hanqi Guo and Zizhong Chen and Franck Cappello},
  title        = {Error-Controlled Lossy Compression Optimized for High Compression Ratios of Scientific Datasets},
  booktitle    = {2018 {IEEE} International Conference on Big Data},
  year         = {2018},
  organization = {IEEE},
  url = {https://doi.org/10.1109/BigData.2018.8622520}
}

@article{almgren2013nyx,
  title={Nyx: A massively parallel amr code for computational cosmology},
  author={Almgren, Ann S and Bell, John B and Lijewski, Mike J and Luki{\'c}, Zarija and Van Andel, Ethan},
  journal={The Astrophysical Journal},
  volume={765},
  number={1},
  pages={39},
  year={2013},
  publisher={The American Astronomical Society},
  url={https://doi.org/10.1088/0004-637X/765/1/39}
}

@article{lukic2015lyman,
  title={The Lyman $\alpha$ forest in optically thin hydrodynamical simulations},
  author={Luki{\'c}, Zarija and Stark, Casey W and Nugent, Peter and White, Martin and Meiksin, Avery A and Almgren, Ann},
  journal={Monthly Notices of the Royal Astronomical Society},
  volume={446},
  number={4},
  pages={3697--3724},
  year={2015},
  publisher={Oxford University Press},
  url={https://doi.org/10.1093/mnras/stu2377}
}

@article{sexton2021nyx,
  title={Nyx: A massively parallel amr code for computational cosmology},
  author={Sexton, Jean and Lukic, Zarija and Almgren, Ann and Daley, Chris and Friesen, Brian and Myers, Andrew and Zhang, Weiqun},
  journal={The Journal of Open Source Software},
  volume={6},
  number={63},
  pages={3068},
  year={2021},
  url={https://doi.org/10.21105/joss.03068}
}

@misc{zstandard,
    series =    {Request for Comments},
    number =    8878,
    howpublished =  {RFC 8878},
    publisher = {RFC Editor},
    doi =       {10.17487/RFC8878},
    url =       {https://doi.org/10.17487/RFC8878},
    author =    {Yann Collet and Murray Kucherawy},
    title =     {{Zstandard Compression and the 'application/zstd' Media Type}},
    pagetotal = 45,
    year =      2021,
    month =     feb,
}

@inproceedings{ae-sz,
  title        = {Exploring Autoencoder-based Error-bounded Compression for Scientific Data},
  author       = {Liu, Jinyang and Di, Sheng and Zhao, Kai and Jin, Sian and Tao, Dingwen and Liang, Xin and Chen, Zizhong and Cappello, Franck},
  booktitle    = {2021 IEEE International Conference on Cluster Computing (CLUSTER)},
  pages        = {294--306},
  year         = {2021},
  organization = {IEEE},
  url={https://doi.org/10.1109/Cluster48925.2021.00034}
}

@inproceedings{SRNN-SZ,
  title        = {Scientific Error-bounded Lossy Compression with Super-resolution Neural Networks},
  author       = {Liu, Jinyang and Di, Sheng and Jin, Sian and Zhao, Kai and Liang, Xin and Chen, Zizhong and Cappello, Franck},
  booktitle    = {2023 IEEE International Conference on Big Data (BigData)},
  pages        = {229--236},
  year         = {2023},
  organization = {IEEE},
  url={https://doi.org/10.1109/BigData59044.2023.10386682}
}

@inproceedings{jia2024gwlz,
  title     = {GWLZ: A Group-wise Learning-based Lossy Compression Framework for Scientific Data},
  author    = {Jia, Wenqi and Jin, Sian and Wang, Jinzhen and Niu, Wei and Tao, Dingwen and Yin, Miao},
  booktitle = {Proceedings of the 14th Workshop on AI and Scientific Computing at Scale using Flexible Computing Infrastructures},
  pages     = {34--41},
  year      = {2024},
  url={https://doi.org/10.1145/3659995.3660041}
}

@ARTICLE{DCINR,
  author={Han, Jun and Yang, Fan},
  journal={IEEE Transactions on Visualization and Computer Graphics}, 
  title={DCINR: A Divide-and-Conquer Implicit Neural Representation for Compressing Time-Varying Volumetric Data in Hours}, 
  year={2025},
  volume={31},
  number={10},
  pages={8116-8128},
  doi={10.1109/TVCG.2025.3564255},
  url={https://doi.org/10.1109/TVCG.2025.3564255}}

@INPROCEEDINGS{sz3simd,
  author={Zou, Changfeng and Ghimire, Bigyan and Calhoun, Jon C.},
  booktitle={2025 IEEE International Conference on Big Data (BigData)}, 
  title={SZ3\_SIMD: Accelerating Error-Bounded Lossy Compression with Architecture Independent SIMD}, 
  year={2025},
  volume={},
  number={},
  pages={4845-4854},
  doi={10.1109/BigData66926.2025.11401598},
  url={https://doi.org/10.1109/BigData66926.2025.11401598}}

@INPROCEEDINGS{cusz+,
  author={Tian, Jiannan and Di, Sheng and Yu, Xiaodong and Rivera, Cody and Zhao, Kai and Jin, Sian and Feng, Yunhe and Liang, Xin and Tao, Dingwen and Cappello, Franck},
  booktitle={2021 IEEE International Conference on Cluster Computing (CLUSTER)}, 
  title={Optimizing Error-Bounded Lossy Compression for Scientific Data on GPUs}, 
  year={2021},
  volume={},
  number={},
  pages={283-293},
  doi={10.1109/Cluster48925.2021.00047}}

@inproceedings{cuszhi,
author = {Wu, Shixun and Pan, Jinwen and Liu, Jinyang and Tian, Jiannan and Qiu, Ziwei and Huang, Jiajun and Zhao, Kai and Liang, Xin and Di, Sheng and Chen, Zizhong and Cappello, Franck},
title = {Boosting Scientific Error-Bounded Lossy Compression through Optimized Synergistic Lossy-Lossless Orchestration},
year = {2025},
isbn = {9798400714665},
publisher = {Association for Computing Machinery},
address = {New York, NY, USA},
url = {https://doi.org/10.1145/3712285.3759798},
doi = {10.1145/3712285.3759798},
booktitle = {Proceedings of the International Conference for High Performance Computing, Networking, Storage and Analysis},
pages = {2024–2037},
numpages = {14},
location = {
},
series = {SC '25}
}

@inproceedings{cuszp3,
author = {Huang, Yafan and Di, Sheng and Li, Guanpeng and Cappello, Franck},
title = {GPU Lossy Compression for HPC Can Be Versatile and Ultra-Fast},
year = {2025},
isbn = {9798400714665},
publisher = {Association for Computing Machinery},
address = {New York, NY, USA},
url = {https://doi.org/10.1145/3712285.3759817},
doi = {10.1145/3712285.3759817},
booktitle = {Proceedings of the International Conference for High Performance Computing, Networking, Storage and Analysis},
pages = {2056–2075},
numpages = {20},
location = {
},
series = {SC '25}
}

@article{z-checker,
  author  = {Dingwen Tao and Sheng Di and Hanqi Guo and Zizhong Chen and Franck Cappello},
  doi     = {10.1177/1094342017737147},
  journal = {The International Journal of High Performance Computing Applications},
  number  = {2},
  pages   = {285--303},
  title   = {Z-checker: A framework for assessing lossy compression of scientific data},
  volume  = {33},
  year    = {2019}
}

@article{jhtdb,
  title={A public turbulence database cluster and applications to study Lagrangian evolution of velocity increments in turbulence},
  author={Li, Yi and Perlman, Eric and Wan, Minping and Yang, Yunke and Meneveau, Charles and Burns, Randal and Chen, Shiyi and Szalay, Alexander and Eyink, Gregory},
  journal={Journal of Turbulence},
  number={9},
  pages={N31},
  year={2008},
  publisher={Taylor \& Francis},
  url={https://doi.org/10.1080/14685240802376389}
}

@misc{nyx,
  author       = {{NYX simulation}},
  howpublished = {\url{https://amrex-astro.github.io/Nyx}},
  note         = {Online},
  year         = {2019}
}

@ONLINE{miranda,
title={Miranda application},
url = {https://wci.llnl.gov/simulation/computer-codes/miranda},
}

@misc{hurricane,
  author       = {{Hurricane ISABEL simulation data}},
  howpublished = {\url{http://vis.computer.org/vis2004contest/data.html}},
  note         = {Online},
  year         = {2004}
}

@inproceedings{sanjaysc25,
author = {Li, Xiao and Zhu, Liangji and Rangarajan, Anand and Ranka, Sanjay},
title = {Generative Latent Diffusion for Efficient Spatiotemporal Data Reduction},
year = {2025},
isbn = {9798400714665},
publisher = {Association for Computing Machinery},
address = {New York, NY, USA},
url = {https://doi.org/10.1145/3712285.3759836},
doi = {10.1145/3712285.3759836},
booktitle = {Proceedings of the International Conference for High Performance Computing, Networking, Storage and Analysis},
pages = {1980–1991},
numpages = {12},
location = {
},
series = {SC '25}
}

@inproceedings{zhang2023fz,
  title={Fz-gpu: A fast and high-ratio lossy compressor for scientific computing applications on gpus},
  author={Zhang, Boyuan and Tian, Jiannan and Di, Sheng and Yu, Xiaodong and Feng, Yunhe and Liang, Xin and Tao, Dingwen and Cappello, Franck},
  booktitle={Proceedings of the 32nd International Symposium on High-Performance Parallel and Distributed Computing},
  pages={129--142},
  year={2023}
}

@ARTICLE{fpzip,
  author={Lindstrom, Peter and Isenburg, Martin},
  journal={IEEE Transactions on Visualization and Computer Graphics}, 
  title={Fast and Efficient Compression of Floating-Point Data}, 
  year={2006},
  volume={12},
  number={5},
  pages={1245-1250},
  doi={10.1109/TVCG.2006.143}}
\end{document}